%% file: arxiv_sept.tex
\documentclass[a4paper,fleqn,usenatbib]{mnras}

\usepackage[T1]{fontenc}
\usepackage{ae,aecompl}

\usepackage{graphicx}	
\usepackage{amsmath}	
\usepackage{amssymb}	%
\usepackage{deluxetable}
\usepackage{xspace}
\usepackage{pdflscape}

\bibpunct{(}{)}{;}{a}{}{,}
\usepackage{hyperref}
\hypersetup{
    colorlinks,
    citecolor=blue,
    filecolor=blue,
    linkcolor=blue,
    urlcolor=blue
}

\title[Satellites around Dwarfs]{An observer's guide to the (Local Group) dwarf galaxies: predictions for their own dwarf satellite populations}

\author[G. A. Dooley et al.]{\parbox{17cm}{
Gregory A. Dooley$^{1}$\thanks{e-mail: greg.dooley@gmail.com},
Annika H. G. Peter$^{2,3}$,
Tianyi Yang$^{4}$,
Beth Willman$^{5}$,
Brendan F. Griffen$^{1}$
and Anna Frebel$^{1}$,
}\vspace{0.3cm}\\
$^{1}$Department of Physics and Kavli Institute for Astrophysics and Space Research, Massachusetts Institute of Technology, \\ Cambridge, MA 02139, USA\\
$^{2}$CCAPP and Department of Physics, The Ohio State University, Columbus, OH 43210, USA \\
$^{3}$Department of Astronomy, The Ohio State University, Columbus OH 43210, USA\\
$^{4}$Institute of Optics, University of Rochester, Rochester, New York, 14627, USA\\
$^{5}$Steward Observatory and LSST, 933 North Cherry Avenue, Tucson, AZ 85721, USA\\
}

\date{Accepted by \mnras \ 2017 July 22. Received 2017 July 22; in original 2016 September 27} 
\pubyear{2017}

\newcommand{\msun}{$\, \mathrm{M_{\sun}}$\xspace}
\newcommand{\lsun}{$\, \mathrm{L_{\sun}}$\xspace}
\newcommand{\rvir}{$R_{\mathrm{vir}}$\xspace}
\newcommand{\mvir}{$M_{\mathrm{vir}}$\xspace}
\newcommand{\mhost}{$M_{\mathrm{host}}$\xspace}
\newcommand{\msub}{$M_{\mathrm{sub}}$\xspace}

\newcommand{\mtwo}{$M_{\mathrm{200}}$\xspace}

\newcommand{\mhalo}{$M_{\mathrm{halo}}$\xspace}
\newcommand{\minftwo}{$M_{\rm{200}}^{\rm{infall}}$\xspace}
\newcommand{\mpeakthree}{$M_{\rm{350}}^{\rm{peak}}$\xspace}
\newcommand{\mpeakvir}{$M_{\rm{vir}}^{\rm{peak}}$\xspace}

\newcommand{\mstarmhalo}{$M_* - M_{\rm{halo}}$\xspace}
\newcommand{\mstar}{$M_*$\xspace}
\newcommand{\lstar}{$L_*$\xspace}
\newcommand{\nbarlum}{$\bar{N}_{\rm{lum}}$\xspace}
\newcommand{\nbarlumfive}{$\bar{N}_{\rm{lum}}^5$\xspace}
\newcommand{\vmax}{$v_{\rm{max}}$\xspace}
\newcommand{\vmaxpre}{$v_{\rm{max}}^{\rm{pre}}$\xspace}
\newcommand{\vmaxfilt}{$v_{\rm{max}}^{\rm{filt}}$\xspace}

\begin{document}
\label{firstpage}
\pagerange{\pageref{firstpage}--\pageref{lastpage}}
\maketitle

\begin{abstract}  
A recent surge in the discovery of new ultrafaint dwarf satellites of the Milky Way has inspired the idea of searching for faint satellites, $10^3$\msun$< M_* < 10^6$\msun, around less massive field galaxies in the Local Group. Such satellites would be subject to weaker environmental influences than Milky Way satellites, and could lead to new insights on low mass galaxy formation. In this paper, we predict the number of luminous satellites expected around field dwarf galaxies by applying several abundance matching models and a reionization model to the dark-matter only \textit{Caterpillar} simulation suite. For three of the four abundance matching models used, we find a $>99\%$ chance that at least one satellite with stellar mass $M_*> 10^5$\msun exists around the combined five Local Group field dwarf galaxies with the largest stellar mass. When considering satellites with $M_*> 10^4$\msun, we predict a combined $5-25$ satellites for the five largest field dwarfs, and $10-50$ for the whole Local Group field dwarf population. Because of the relatively small number of predicted dwarfs, and their extended spatial distribution, a large fraction each Local Group dwarf's virial volume will need to be surveyed to guarantee discoveries. We compute the predicted number of satellites in a given field of view of specific Local Group galaxies, as a function of minimum satellite luminosity, and explicitly obtain such values for the Solitary Local dwarfs survey. Uncertainties in abundance matching and reionization models are large, implying that comprehensive searches could lead to refinements of both models.

\end{abstract}


\section{Introduction}
Hierarchical structure formation in the Lambda Cold Dark Matter Universe predicts that galaxies like the Milky Way (MW) and M31 are orbited by satellite galaxies \citep{Frenk12}. Observations have long supported this hierarchical accretion model, starting with identifying that the Large Magellanic Cloud (LMC) and Small Magellanic Cloud (SMC) are within close proximity of the Milky Way \citep{Shapley22, Shapley24}. An additional nine MW satellites with luminosity $L_* > 10^5$\lsun, the classical dwarfs, were discovered next, followed by a class of satellite galaxies with luminosity $10^3 < L < 10^5$\lsun, the ultrafaint dwarfs (UFDs), initially discovered in the Sloan Digital Sky Survey \citep{Willman05b,Zucker06a,Belokurov06,Belokurov07,belokurov2008,belokurov2010,Irwin07,Walsh07}. Even smaller ``hyperfaint'' galaxies ($L_* < 10^3$\lsun) have also been discovered, galaxies so tiny they can only be found very near the Sun \citep{Willman05a,Zucker06b,belokurov2009}. 

The window into UFD and hyperfaint satellites of the MW is opening up dramatically, as the Dark Energy Survey, PanSTARRS, ATLAS, and MagLiteS surveys have found $\sim$20 new UFD satellite candidates in the past two years (\citealt{Bechtol15}; \citealt{Drlica15}; \citealt{Kim15a}; \citealt{Kim15b}; \citealt{Koposov15}; \citealt{Laevens15}; \citealt{Martin15}; \citealt{Luque16}; \citealt{Torrealba16a}; \citealt{Drlica-Wagner16}), and will likely continue to find more.

These recent discoveries, along with follow up observations, have opened the door to better understand low mass galaxy formation. Whereas classical dwarfs have recent star formation, many UFDs have been confirmed to be ``fossil'' galaxies, meaning that $>70\%$ of their stars formed before reionization \citep{Brown12, Brown14a, Brown14b}. Consequently, they contain very old stellar populations \citep{Kirby08, Norris10, Frebel12} and are ideal targets to learn about early universe galaxy formation.  Both classical dwarfs and UFDs serve as probes on the interplay of ionizing radiation, supernova feedback, star formation, and halo size. Low-mass galaxies are susceptible to losing gas from reionization and supernovae, which can turn them into fossil galaxies or even leave them entirely dark. However, due to strong environmental effects on dwarf galaxy evolution, and halo-to-halo variation, a large sample size of dwarf galaxies in different environments is necessary to probe the halo size scale where these effects begin.

To constrain star formation models in dwarf galaxies, there is significant value in even just the number counts of UFDs and classical dwarfs. Completeness in discovery around the MW will further refine the so-called ``missing satellite problem'' \citep{Moore99, Klypin99} and its many proposed solutions, which come in both baryonic and dark-matter flavors.  Baryonic solutions include the effects described above, to suppress the formation of stars in small dark-matter halos.  With a large sample of UFDs, we can test abundance-matching-derived \mstarmhalo relationships to much smaller mass scales than those for which the relations were observationally inferred.  A change in the slope or scatter of abundance-matching relations would have significant implications for the drivers of star-formation efficiency in small galaxies.

Discovering a large sample of UFDs is also critical in revealing a diversity of chemical enrichment pathways, such as r-process enhancement or lack thereof \citep{Ji16}. By finding more UFDs in particular, we should discover older and more metal poor stellar populations. We also create more opportunities to measure internal halo structure, which has important implications on the cusp/core debate \citep{deBlok10} and the nature of dark matter \citep{Elbert15,Dooley16}.
 
Given the importance of finding more low-mass galaxies, it is natural to consider searching for them beyond the MW. Already, many satellites have been discovered around Andromeda \citep{Zucker04, Zucker07,McConnachie08, McConnachie09, Majewski07, Irwin08, Martin09, Bell11, Slater11, Richardson11}, and around a handful of nearby galaxies and clusters \citep{Jang14,sand2014,Crnojevic16}.  Several of these recent discoveries have been of dwarf galaxy satellites of dwarf galaxies themselves \citep[e.g.]{sand2015,Carlin16}. This opens the question if there could exist satellites around isolated dwarf galaxies within the Local Group itself. Due to a lower mass host, they would experience weaker environmental influences than those in the MW or M31. Tidal and ram pressure stripping are reduced, so satellites would retain more of their original stars, gas, and dark matter. Reionization would also proceed differently, as the nature of the closest source of ionizing photons would change. These differences would provide an opportunity to better isolate the internal drivers of low mass galaxy formation. While isolated galaxies would have even weaker environmental effects, hierarchical galaxy formation dictates that the density of low-mass galaxies is greater around a larger galaxy than in areas of complete isolation. There are already some hints that the relationship between stellar mass and halo mass is different in dense environments compared to more field-like environments \citep{Grossauer15,Read17}.  Studying a sample of dwarfs, spanning the range from UFD to more recently star-forming and larger dwarfs ($M_* \gtrsim 10^5$\msun), in less-dense environments is important to determining what physics shapes the relation between dark matter halos and the galaxies they host.

We therefore set out to characterize the abundance of satellites around Local Group field dwarfs, or ``dwarf-of-dwarf", systems as a guide to current and future surveys. We predict the number of satellites of dwarf galaxies given simple, physically motivated prescriptions for how dwarfs populate dark-matter halos in the canonical cold-dark-matter model.  We outline observational strategies for finding dwarf-of-dwarf satellites and discuss how to interpret observations in light of models for star formation in small halos. We focus specifically on the satellite systems of Local Group field dwarf galaxies, because the proximity of these galaxies enables the discovery of very low luminosity satellites as overdensities of resolved stars. We include specific predictions for the fields of view of the Solitary Local dwarfs survey (Solo), a recent survey of all isolated dwarfs within $3$ Mpc of the Milky Way \citep{Higgs16}. Though the main goals of the Solo survey do not include finding satellites, it likely already has at least one lurking in its data. Furthermore, our results can be used to estimate the number of dwarf-of-dwarf satellites which Sagittarius, Fornax, and the SMC brought into the Milky Way at infall. Values for the LMC, which is larger than the mass range of hosts we consider in this paper, will be presented in future work.

Using semianalytic models and hydrodynamic simulations, \cite{Sales13} and \cite{Wheeler15} have made similar calculations to ours, predicting the probability of an UFD satellite around a dwarf galaxy within a $10^{10}$\msun dark matter halo, as $40-50\%$ and $35\%$, respectively. We perform a more in-depth study over a larger parameter space, finding the likelihood of satellites existing around dwarf galaxy hosts of a range of host masses, the mean number of satellites around hosts as a function of satellite stellar mass, and the full probability distribution of the number of satellites around known field dwarfs. Due to uncertainty in the \mstarmhalo relationship for low luminosity systems, we use a variety of abundance matching models rather than just one model for star formation. We additionally determine the sensitivity of predictions on input parameters, including reionization, a study not previously conducted.

Our paper is organized as follows:
In Section~\ref{sec:methods}, we outline our methods for modeling satellite populations in isolated dwarf galaxies. In Section~\ref{sec:results}, we validate our methods with predictions for the Milky Way, predict how many luminous satellites should exist around dwarf galaxies, and compute the probability of finding one or more satellites per host. In Section~\ref{sec:radial}, we provide a model for the number of satellites within a line of sight as a function of field of view, and comment on observational strategies. In Section~\ref{sec:sensitivity}, we show how sensitive our predictions are to uncertainties. Finally, we summarize our key findings and present a plan for future directions in Section~\ref{sec:conclusions}.

\section{Methods}
\label{sec:methods}
To predict the number of luminous satellites of Local Group dwarf galaxies, we apply a suite of abundance matching models and a parameterized reionization recipe to dark-matter-only simulations of Milky Way-like halos (and their surrounding environments).  This simple scheme is fast to implement, unlike fully hydrodynamic simulations, and allows us to quickly explore different models for how dwarf galaxies populate halos.  With this scheme, we can generate many realizations of dwarf satellite systems, so we can define a probability distribution for the satellite populations for each model.

We use dark-matter only simulations to predict the subhalo mass functions (SHMFs) of satellites around isolated field dwarfs in the vicinity of a Milky Way-mass galaxy, and use these SHFMs to generate Poisson samples of subhalos around each dwarf galaxy host. Next, we model the effects of reionization by assigning each subhalo a probability that it hosts stars or remains dark. We then apply abundance matching prescriptions from the literature to assign stellar mass to the luminous subhalos. 

In the following subsections we elaborate on the simulations, abundance matching models, reionization methodology, and mass functions used.

\subsection{\textit{Caterpillar} Simulation Suite}
\label{sec:simulations}
We use a sample of $33$ high particle resolution ($m_{\rm{p}} = 3\times 10^4$\msun) and high temporal resolution ($320$ snapshots) zoom-in simulations of Milky Way-sized galaxies from the \textit{Caterpillar} simulation suite \citep{Griffen16}. The simulations are used to determine the typical SHMF, radial dependence of the SHMF, subhalo infall distribution times, and dark fraction of halos due to reionization. We perform these calculations on both the Milky Way sized host halo and smaller nearby field halos. We consider field galaxies as halos with virial mass between $10^{10}$ and $10^{11.5}$\msun at $z=0$ that are outside of the virial radius of the MW sized host, and within the uncontaminated volume of each simulation. We choose the mass range to reflect that of real Local Group field galaxies. In total there are $148$ field halos across the $33$ simulations. 

All self-bound haloes are found using a modified version of the {\sc{rockstar halo finder}} \citep{Behroozi13RS} which includes full iterative unbinding to improve halo finding accuracy, as described in \cite{Griffen16}. Merger trees were produced by {\sc{rockstar consistent trees}} \citep{Behroozi13MT}. Any mention of \textit{virial} refers to the \cite{Bryan98} definition of the virial radius, \rvir, which at $z=0$ for our cosmological parameters is the radius such that the mean enclosed halo density is $104$ times the critical density of the universe, $\rho_c = 3H_0^2/8 \rm{\pi} G$. \mvir refers to the gravitationally bound mass within \rvir, and any mention of $R_{\mathrm{\Delta}}$ or $M_{\mathrm{\Delta}}$ refers to the radius and mass of a halo where the mean enclosed density is $\rm{\Delta}$ times the critical density.

\subsection{Abundance Matching Models}
\label{sec:abundance_matching}
\begin{figure}
\includegraphics[width=0.48\textwidth]{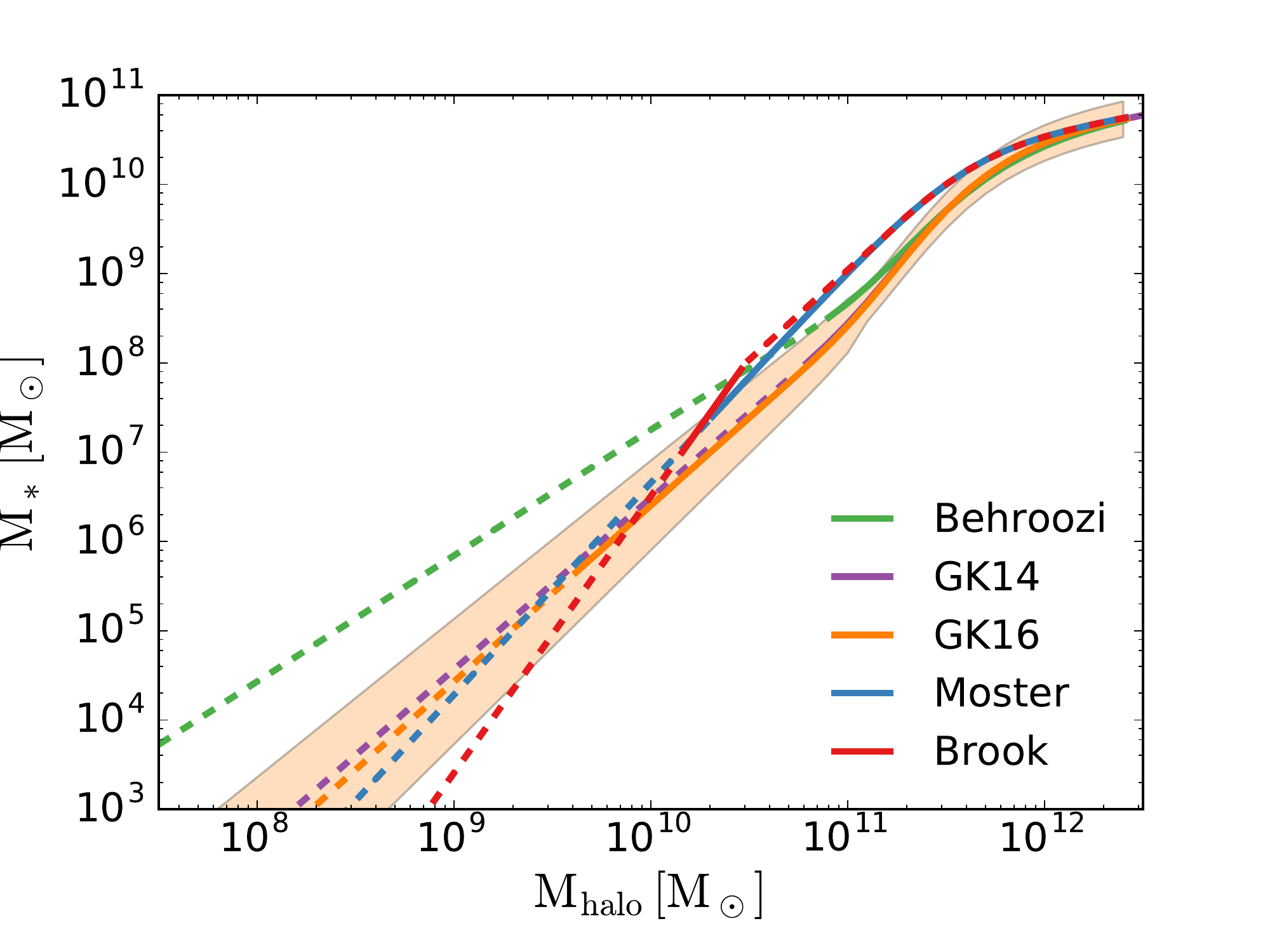}
\caption[Abundance matching derived stellar mass-halo mass relationships]{Abundance matching derived stellar mass-halo mass relationships for several recent models. Solid lines indicate ranges of each model where they were matched to observations, dashed lines indicate ranges of extrapolation. The large variation in predictions, particularly for halos with \mhalo $< 10^{10}$\msun, results in very different predictions for the number of low mass satellite galaxies that could be discovered in the Local Group. While similar, the definition of \mhalo is different for each of the models, making purely visual comparisons between functions not entirely accurate. The mass definitions are listed in Table~\ref{table:shmf_params}. The shaded area around GK16 indicates the $\pm 1 \sigma$ lognormal scatter we implement.}
\label{fig:AMrelationships}
\end{figure}

Abundance matching (AM) is a technique employed to determine an approximate stellar mass to halo mass (\mstarmhalo) relationship for galaxies. Given a set of observed galaxies within a volume down to some luminosity completeness limit, galaxies are matched in a one-to-one fashion with dark matter halos from a simulation of the same volume. They traditionally assume a monotonic relationship of stellar mass and dark matter halo mass to create a function $M_*($\mhalo$)$ that satisfies the condition
\begin{equation} \label{eq:AM}
\int_{M_*(m_1)}^{M_*(m_2)} \frac{\rm{d}N_*}{\rm{d}M}(M_*) \rm{d}M_* = \int_{m_1}^{m_2} \frac{\rm{d}N}{\rm{d}M}(M_{halo}) \rm{d}M_{halo} 
\end{equation}
where $\frac{\rm{d}N_*}{\rm{d}M}(M_*)$ is the differential stellar mass function, and $\frac{\rm{d}N}{\rm{d}M}$ is the differential halo mass function \citep{Yang03,Vale04, Kravtsov04,Tasitsiomi04,Vale06, Guo10, Moster10,Kravtsov10,Wang10, Yang12, Moster13,Behroozi13AM, Brook14,GarrisonKimmel14}.  
For galaxies with $M_* > 10^8$\msun, abundance matching relationships produce relatively consistent results with each other. However, at smaller masses, incomplete surveys of low luminosity galaxies and a more stochastic process of star formation in halos leads to larger uncertainty in the \mstarmhalo relationship. We highlight this in Fig.~\ref{fig:AMrelationships}, showing the relationship for several recently proposed models. Different extrapolations of the function down to low masses, how stochastic star formation is, and what simulations and observations were compared lead to very different predictions. We briefly describe each of the models and details on their implementation in the following paragraphs. The names in bold indicate how we refer to the models in the rest of the paper.

\begin{figure}
\includegraphics[width=0.48\textwidth]{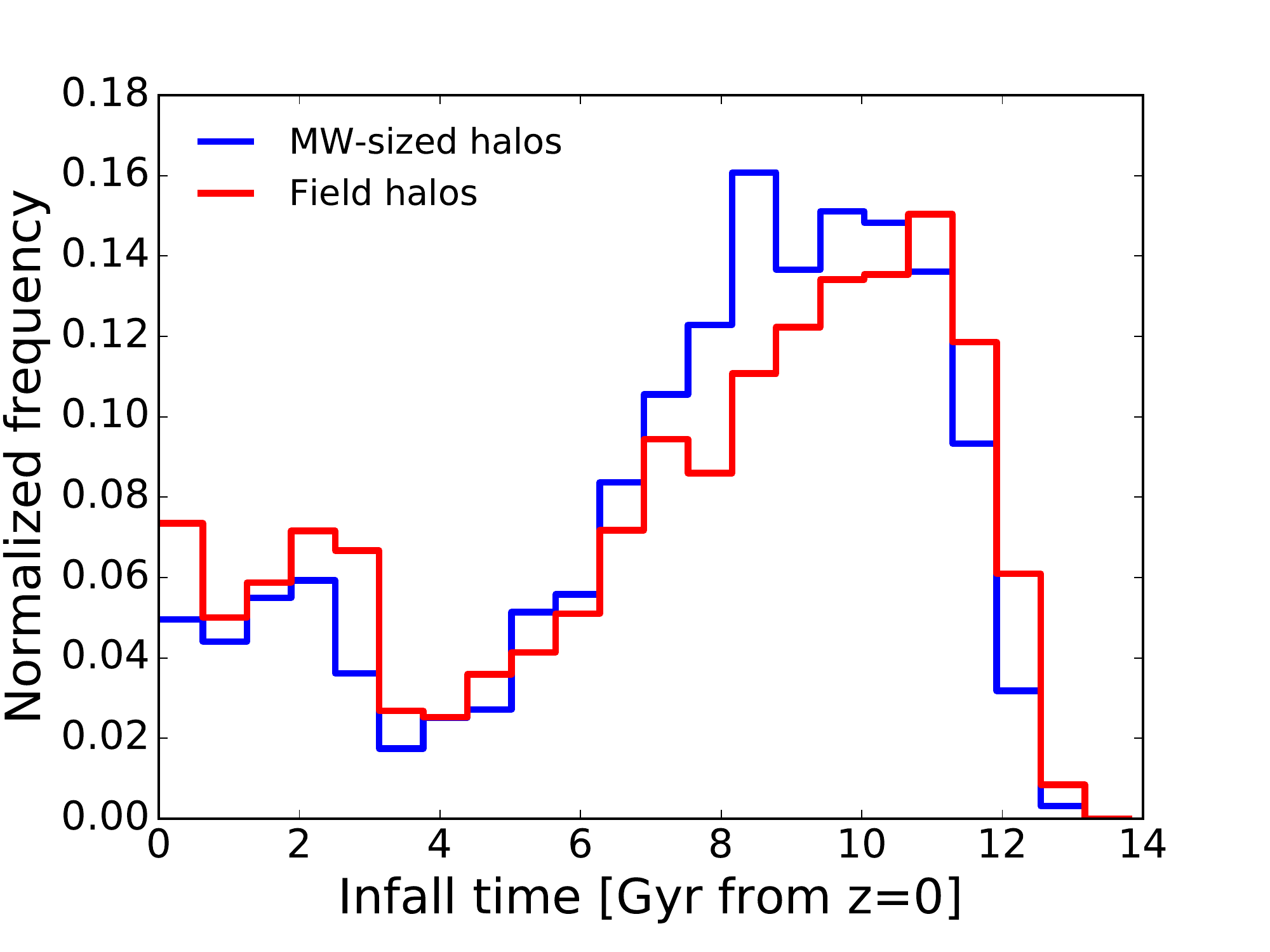} 
\caption[Infall time distribution of $z=0$ subhalos]{Infall time distribution for all $z=0$ subhalos from 33 \textit{Caterpillar} simulations that are part of MW-sized hosts (blue) and part of field halos (red). Time is given by the duration between infall and $z=0$. The distribution does not vary significantly with host halo mass, nor range of subhalo mass.}
\label{fig:infall}
\end{figure}

\textbf{Moster:} \cite{Moster13} match observed stellar mass functions at different redshifts from Sloan Digital Sky Survey (SDSS), Spitzer Space Telescope, Hubble Space Telescope and Very Large Telescope to dark matter halos in the Millennium \citep{Springel05} and Millennium-II \citep{Boylan-Kolchin09} simulations to produce a redshift dependent AM model. For subhalos, they define \mhalo as \minftwo, the mass of a halo at first infall enclosed by a volume that is $200 \times \rho_c$. To account for the redshift dependence in their model, we find the infall time distributions of all $z=0$ satellites in \textit{Caterpillar} MW-sized hosts and field halos, and use them to assign random infall times to our subhalos in subsequent analysis. We investigate whether the infall time distribution changes when considering different ranges of subhalo masses, but find at most a weak dependence on subhalo mass that results in a $<1\%$ influence on our final estimates of luminous satellites. We therefore use an infall time distribution independent of subhalo mass. We find the distributions for satellites in MW-sized halos and field halos are consistent with each other, as seen in Fig.~\ref{fig:infall}, and match well in form to fig. 3 of \cite{Barber14}. \cite{Barber14} finds that when selecting only subhalos which form stars as opposed to all subhalos, the mean infall time is shifted $\sim 1$ Gyr earlier. This adjustment makes little difference to the predictions made by the Moster model, as discussed in Section~\ref{sec:sensitivity}, so we use the distribution for all subhalos, taken directly from the data as binned in Fig.~\ref{fig:infall}.

\textbf{Behroozi:} \cite{Behroozi13AM} deduce stellar mass functions from $z=0$ to $z=8$ with results from SDSS, \textit{GALEX}, and PRIMUS surveys. They match these to halo mass functions from the \textit{Bolshoi} \citep{Klypin11}, \textit{MultiDark} \citep{Riebe11}, and \textit{Consuelo} simulations to produce a redshift dependent \mstarmhalo model which defines \mhalo as \mpeakvir, the maximal virial mass achieved by the subhalo over its history. We show the function in Fig.~\ref{fig:AMrelationships}, but do not include the model in our results since it overpredicts the abundance of low-mass galaxies and is otherwise incorporated in the next two models.

\textbf{GK14:} \cite{GarrisonKimmel14} match galaxies from the SDSS to dark matter halos in their ELVIS suite to create an AM model of the same functional form as that in \cite{Behroozi13AM}, but with a steeper logarithmic slope on the low mass end. They identify that \cite{Behroozi13AM} overestimates the number of galaxies with $M_* < 10^{8.5}$\msun at $z=0$ due to using a now outdated stellar mass function, and correct for it. Below the completeness limit of $M_* = 10^8$\msun, GK14 extrapolates their relationship with a constant slope. Like Behroozi, they define \mhalo as \mpeakvir. For their cosmology, \rvir encloses a volume that has density $97 \times $\rvir, making it marginally larger than $R_{\rm{vir}} = R_{\rm{104}}$ in our cosmology. We find the discrepancy small enough to not take into account in detail.

\textbf{GK16:} Many hydrodynamic simulations have demonstrated that there can be significant scatter about a mean \mstarmhalo relationship \citep{Munshi13, Sawala15, Wheeler15, OShea15}. The scatter increases towards lower masses, making the default abundance matching assumption of a monotonic relationship problematic particularly for dwarfs \citep{Power14, Ural15}. \cite{GarrisonKimmel16} explicitly model the scatter, proposing a range of stochastic abundance matching relationships. They build off the GK14 model, changing the slope of the best fit relationship as a function of the 1-sigma level of lognormal scatter, $\sigma_{\rm{scat}}$. With higher levels of scatter, more galaxies are upscattered above the observed completeness luminosity than are scattered below it due to the increasing abundance of DM subhalos at lower masses. Consequently, their mean \mstarmhalo relationship must be steeper and thereby lower for low mass halos. They match galaxies in the Local Group to the ELVIS suite down to a completeness limit of $M_* > 4.5 \times 10^5$\msun, and define \mhalo as \mpeakvir as in GK14.

Our default implementation of this model is to use the ``growing scatter'' model, in which $\sigma_{\rm{scat}}$ grows for decreasing halo masses. We make this choice because simulations such as those of \cite{OShea15} and \cite{Sawala15} support a growing scatter more than a constant scatter. The level of growth is dictated by a parameter, $\gamma$, as in Eq. (3) of \cite{GarrisonKimmel16}. We choose a default value of $\gamma = -0.2$. In Section~\ref{sec:sensitivity}, we discuss how results change when varying $\gamma$. We implement the scatter by sampling a lognormal offset from the mean \mstarmhalo relationship randomly from a Gaussian of width $\sigma_{\rm{scat}}(M_{\rm{halo}})$ for each subhalo considered.

\textbf{Brook14:} \cite{Brook14} proposes an even steeper slope than \cite{GarrisonKimmel14}, which, when extrapolated to $M_* < 10^7$\msun, estimates lower stellar masses for a fixed DM mass. They match observed galaxies in the Local Group to the CLUES simulation suite \citep{Gottloeber10}. Instead of \mpeakvir, they define \mhalo as \mpeakthree, the peak mass achieved by a subhalo measured within a volume that has density $350 \times \rho_c$. We implement their model which has an \mstarmhalo log-log slope of $3.1$, and normalization factor $M_0 = 79.6$. For stellar masses $M_* > 10^8$\msun, the Brook model is unspecified, so we linearly interpolate values in log-log space between $M_* = 10^8$\msun in the Brook model and $M_* = 3 \times 10^9$\msun in the Moster model, then switch to values from Moster. No satellites considered in this paper have $M_* > 10^8$\msun, but a few host galaxies do, and a function to estimate their dark matter halo mass from stellar mass is needed.


\subsection{Reionization}
\label{sec:reionization}  
UV photons emitted by the first stars during reionization are able to ionize hydrogen atoms and prevent sufficient cooling and gas accretion needed for star formation \citep{Efstathiou92,Thoul96, Gnedin2000, Wiersma09,Pawlik09}. In low mass halos, they can also heat gas enough to gravitationally escape, sometimes before any star formation begins \citep{Barkana1999, Shapiro04, Okamoto08}. The combination of effects renders many halos entirely dark, an effect recently simulated and emphasized in \cite{Sawala13, Sawala15, Sawala16}. Simply assuming that all dark matter subhalos host luminous galaxies would therefore wildly overestimate the number of visible satellites \citep{Bullock00,Somerville02,Benson02}.

\begin{figure}
\includegraphics[width=0.48\textwidth]{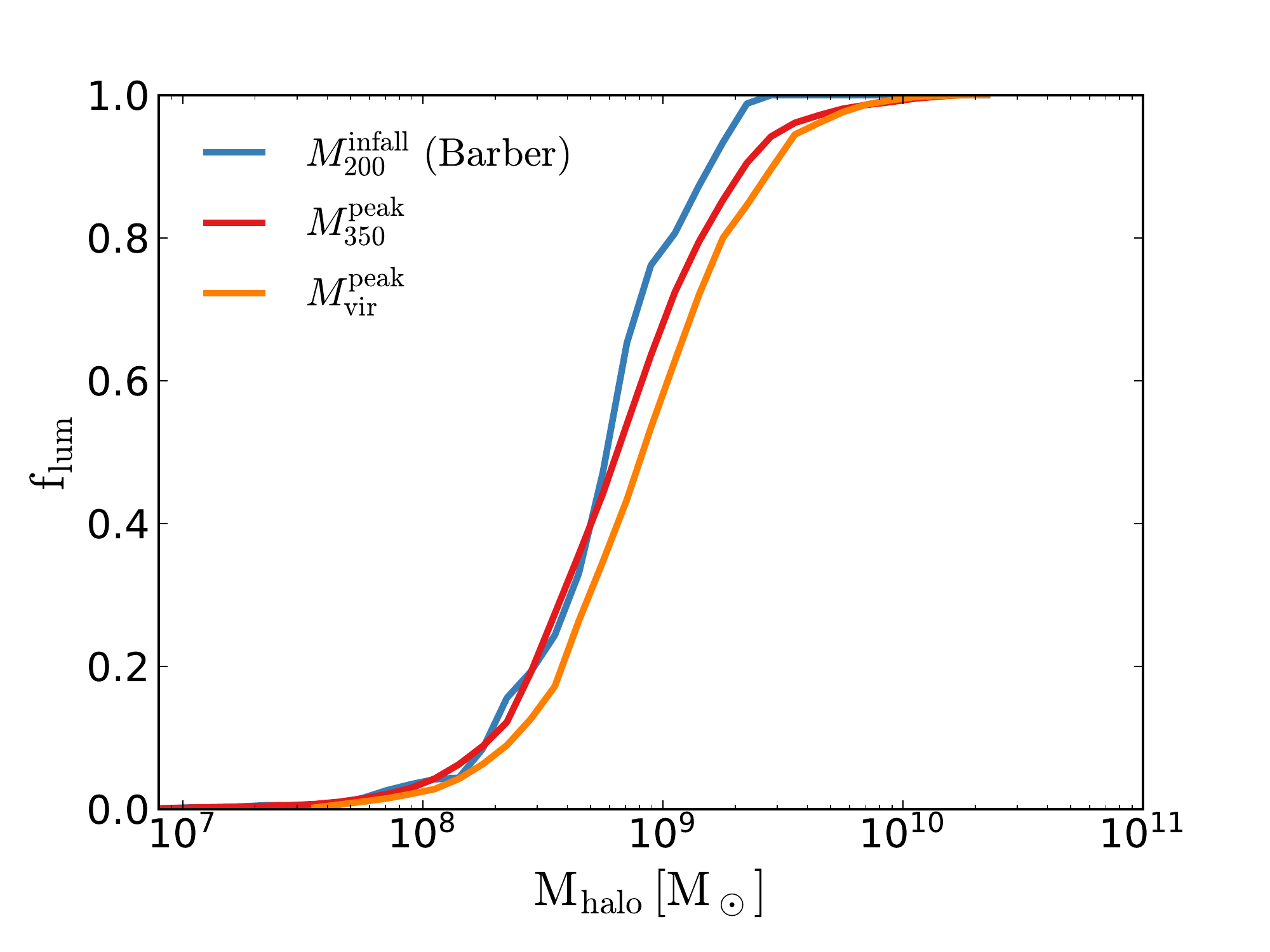}
\caption[Reionization model: fraction of halos that host luminous galaxies]{Fraction of dark matter halos that host luminous galaxies by $z=0$ as a function of different mass definitions. The blue line is smoothed data from \protect\cite{Barber14}, which uses a semi-analytic model applied to dark matter only simulations. The green and red lines are functions for alternate mass definitions, as inferred from the baseline reionization model and \textit{Caterpillar} merger trees.}
\label{fig:lumfrac}
\end{figure}

We model the effects of reionization by randomly assigning halos to host stars or remain dark with probabilities that depend on the halo's mass. Using data obtained from \cite{Barber14}, we produce a smoothed curve indicating the fraction of halos that are luminous at $z=0$ as a function of \minftwo, as plotted in Fig.~\ref{fig:lumfrac}. The function follows from a semi-analytic model applied to the level-2 halos of the Aquarius simulation suite \citep{Springel08}. The model has reionization proceeding from $z=15$ to $z=11.5$. Below a redshift dependent filtering mass, it models photoevaporation by removing baryons from halos. Full details of the semi-analytic model are given in \cite{Starkenburg13}.

Since the abundance matching models use different definitions for \mhalo, we produce a different luminous fraction function for each definition. We do this by randomly assigning halos to be dark or luminous in our simulation according to their \minftwo, then collecting the values of \mpeakvir and \mpeakthree for those same halos in the merger tree. Repeating the random assignments for many instances generates a list of dark and luminous halos paired with each mass definition, which is then turned into the desired function. In all cases, reionization suppresses the number of satellites with $M_* < 10^5$\msun, but has little effect suppressing larger systems.

Many details of reionization, including the redshift of occurrence, environmental effects, $H_{\rm{2}}$ shielding, and the efficiency of photoevaporation, remain uncertain \citep{Onorbe16}, which adds variability to the number of luminous subhalos produced. We therefore investigate alterations to the reionization model and subsequent effects on our results in Section~\ref{sec:sensitivity}.

For that investigation, and for use in determining the radial distribution of satellites which survive reionization, we employ a simple model inspired by \cite{Lunnan12} and \cite{Peter10}. While exploring parameter space with detailed hydrodynamic or semi-analytic reionization models would be more rigorous, it is beyond the scope of this paper. For a halo to form stars by $z=0$, it must either reach a critical size for $\rm{H_2}$ cooling and atomic line cooling before reionization, or become massive enough after reionization to reaccrete and cool gas. We define the size thresholds in terms of \vmax, calling them \vmaxpre and \vmaxfilt respectively. Fixing reionization to happen instantaneously at $z=13.3$, approximately the mean redshift in the model used in \cite{Barber14}, we conduct a parameter search to minimize the difference between the \cite{Barber14} luminous fraction function of Fig.~\ref{fig:lumfrac} and the one produced by applying the \vmax cuts to the merger history in all \textit{Caterpillar} simulations. We achieve a close fit ($< 4\%$ difference at any point) with $v_{\rm{max}}^{\rm{pre}} = 9.5$ and $v_{\rm{max}}^{\rm{filt}} = 23.5 \, \rm{km/s}$. 

Our value for $v_{\rm{max}}^{\rm{pre}}$ is consistent with expectations from the literature. While the threshold for $\rm{H_2}$ cooling is more accurately weakly redshift dependent, it occurs around $M_{\rm{200}} = 10^6$\msun or $T_{\rm{vir}} = 2000-3000 \, \rm{K}$ \citep{Tegmark97, Madau08, Power14}, corresponding to $4-7 \, \rm{km/s}$ at $z=13.3$ in our simulations. Atomic line cooling occurs for larger mass halos, near $M_{\rm{200}} = 10^8$\msun or $T_{\rm{vir}} \approx 4000 \, \rm{K}$ \citep{Bromm11}, corresponding to $16-26 \, \rm{km/s}$ at $z=13.3$ in our simulations. Halos which reach the atomic line cooling limit before reionization nearly universally form and retain stars by $z=0$, whereas $\rm{H_2}$ cooling minihalos may or may not retain stars due to reionization and supernova \citep{Power14}. We thus expect $v_{\rm{max}}^{\rm{pre}}$ to lie between the minimum $\rm{H_2}$ cooling threshold and the atomic cooling threshold, which it does. Moreover, it agrees closely with \cite{Okamoto09} who inferred a value of $v_{\rm{max}}^{\rm{pre}} \approx 12 \, \rm{km/s}$ from hydrodynamic simulations where reionization occurs at $z=8$. If we shift reionization to $z=8$, our best fit value becomes $12.6 \, \rm{km/s}$. 

Our value for $v_{\rm{max}}^{\rm{filt}}$ is consistent with the low end of expectations from the literature. This threshold for star formation to proceed after reionization has been termed the ``filtering mass'', with initial values placed at $20<$ \vmax$<30 \, \rm{km/s}$ \citep{Bovill09, Okamoto09, Bovill11a, Bovill11b}. More recent publications have used higher values of $30<$\vmax$<50 \, \rm{km/s}$ \citep{Peter10,Lunnan12,GarrisonKimmel14, Griffen16b}, highlighting uncertainty in how to model reionization.

\subsection{Mass Functions and Monte Carlo Sampling}
\label{sec:montecarlo}

Using \textit{Caterpillar}, we identify the mean SHMF for all isolated field galaxies and MW analogs. Since the AM models use different mass definitions for \mhalo, we correspondingly find different SHMFs. In each case, the differential number of halos in a given mass interval, $\frac{\rm{d}N}{\rm{d}M_{\rm{sub}}}$, follows the form
\begin{equation}
\label{eq:SHMF}
\frac{\rm{d}N}{\rm{d}M_{\rm{sub}}} = K_0 \left(\frac{M_{\rm{sub}}}{\rm{M_{\sun}}} \right)^{-\alpha} \frac{M_{\rm{host}}}{\rm{M_{\sun}}}
\end{equation}
as has been identified in several previous studies \citep{Gao2004, Bosch2005, Dooley14}. The best fit values of $\alpha$ and $K_0$ do depend weakly on the host halo mass range, but change negligibly within a one dex host mass interval. Since Milky Way-like hosts are more than one dex larger than field halos, we separately compute best fit values of $\alpha$ and $K_0$ for satellites of field halos and satellites of Milky Way-like hosts. We also considered a SHMF form where $\frac{\rm{d}N}{\rm{d}M_{\rm{sub}}}$ is a power law function of $M_{\rm{sub}}/M_{\rm{host}}$ rather than being directly proportional to the host mass, but find the best fit parameters in this case are more sensitive to the host halo mass range considered. The values $\alpha$ and $K_0$ are computed for each mass definition and shown in Table~\ref{table:shmf_params}. The mass functions count all self-bound subhalos within \rvir at $z=0$ regardless of the mass definition. $M_{\rm{host}}$, however, uses the same mass definition as $M_{\rm{sub}}$.

Subhalos of subhalos are excluded to be consistent with the methods of the AM models we use. We nonetheless encourage future models to include sub-subhalos, as we find our mass functions increase by $\sim 40\%$ when including all levels of nested substructure.

\begin{figure}
\includegraphics[width=0.48\textwidth]{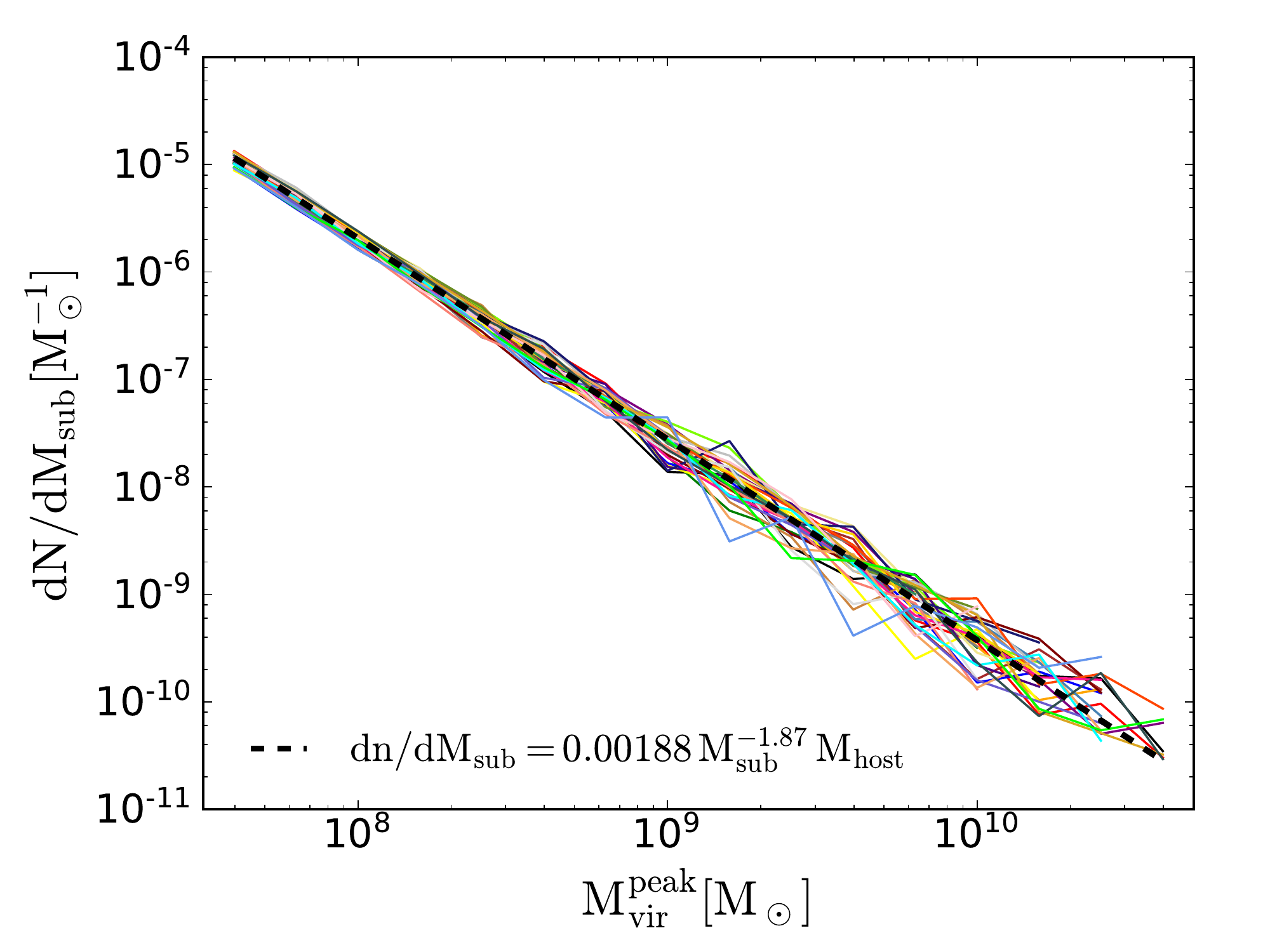}
\caption[Subhalo mass functions]{Subhalo mass functions of $33$ \textit{Caterpillar} simulations, each scaled and plotted as though the host has a mass of \mhost$= 10^{12}$\msun. The scaling is done in accordance with equation~(\ref{eq:SHMF}). The black dashed line indicates the fit to the average of all SHMFs, and is consistent with the written equation. All values of parameters in Table~\ref{table:shmf_params} were obtained in this way. For all fits, a subhalo mass of \msub$= 10^{7.5}$\msun is used as the lower bound, where the SHMFs are still converged.}
\label{fig:shmf}
\end{figure}

A representative SHMF fit is shown in Fig.~\ref{fig:shmf}, using a subhalo mass definition of \mpeakvir. Fits to the other definitions are nearly identical in form. The SHMF for each simulation is scaled to that of a host with mass \mhost$= 10^{12}$\msun before their average is taken and a fit applied. The function is fit down to \msub$ = 10^{7.5}$\msun, a value where the mean stellar mass per halo drops below $10^3$\msun in all AM models used, and where the SHMF is still well converged for all mass definitions.

\begin{table*}
\tablewidth{0.87\textwidth}
\centering
\caption{Subhalo mass functions for MW size halos and dwarf field halos}
\label{table:shmf_params}
\begin{tabular}{cccccc}
\hline
\hline
\textbf{Mass Definition} &  $\pmb{\alpha}$ & $\pmb{K_0}$ & $\pmb{\alpha}$ & $\pmb{K_0}$ & \textbf{AM Model(s)} \\
 & (field) & (field) & (MW analog) & (MW analog) &  \\
\hline
\minftwo & 1.81 & 0.000635 & 1.84 & 0.000854 & Moster \\
\hline
\mpeakvir & 1.82 & 0.000892 & 1.87 & 0.00188 & GK14, GK16, Behroozi \\
\hline
\mpeakthree & 1.81 & 0.000765 & 1.87 & 0.00200 & Brook \\
\hline

\end{tabular}
\tablecomments{Values of $\alpha$ and $K_0$ in equation~(\ref{eq:SHMF}) for the mean subhalo mass function for various definitions of subhalo mass. Columns two and three designate values for isolated dwarf field galaxies, and four and five for Milky Way analogs. The parameters of the mass function are approximately independent of host halo mass over a one dex interval.  Milky Way analogs and field dwarf halos need to be separated, but within each category one set of parameters is sufficient. The abundance matching models employing each definition are indicated.}
\end{table*}

The mean number of dark matter subhalos, $\bar{N}$, around a host of mass $M_{\rm{host}}$ is found by integrating equation~(\ref{eq:SHMF}) from $M_{\rm{min}}$ to $M_{\rm{host}}$ to yield:
\begin{equation}
\label{eq:N}
\bar{N} = \frac{K_0 M_{\rm{host}}}{\alpha -1} \left( M_{\rm{min}}^{1-\alpha} - M_{\rm{host}}^{1-\alpha} \right)
\end{equation}
where $M_{\rm{min}}$ is a halo mass at which no star formation occurs in any model due to reionization. We choose a conservative value of $10^{7.4}$\msun. Since subhalo abundances are approximately Poisson distributed around the mean, we generate random realizations of the number of subhalos between $M_{\rm{min}}$ and $M_{\rm{host}}$ according to a Poisson distribution with a mean $\lambda = \bar{N}$, and then randomly assign halo masses to them according to the SHMF. More accurately, subhalo abundances follow a negative binomial distribution with the variance of $\bar{N}$ increasing relative to that of a Poisson distribution as $M_{\rm{sub}} / M_{\rm{host}}$ decreases \citep{BoylanKolchin10, Mao15, Lu16}. However, \cite{BoylanKolchin10} find that the super Poissonian spread only becomes important for $M_{\rm{sub}} / M_{\rm{host}} \leq 10^{-3}$. Otherwise Poisson statistics remain a good approximation, as verified in the \textit{Caterpillar} suite. The largest field halo we consider is $\sim 6 \times 10^{10}$\msun, and only subhalos with \mvir$ \gtrsim 10^8$\msun contribute to satellites with \mstar$> 10^3$\msun, so the minimum ratio relevant to this study is $M_{\rm{sub}} / M_{\rm{host}} = 1.7 \times 10^{-3}$. In this instance, $\sigma_{\bar{N}} / \sigma_{\rm{Poisson}} \leq 1.2$. For all other field halos and for abundances of more luminous satellites, our Poisson approximation is even more accurate.

Once halo masses are assigned, they are chosen to be luminous or dark with probabilities following the luminous fraction as a function of halo mass as shown in Fig.~\ref{fig:lumfrac}. The luminous halos are then assigned a stellar mass according to the \mstarmhalo relationship and scatter (if any) of the model under consideration.

\subsection{Inferring \mhalo from \mstar}
\label{sec:mstar_to_mhalo}
Given a target host halo, its total virial mass is needed in order to generate a realization of its SHMF. However, the dark matter mass is essentially impossible to measure directly for Local Group field dwarfs.  Lensing methods are not available, and the mapping between the dynamical mass within the half-light radius of the galaxy (with stellar kinematics) to the virial mass is highly dependent on the dark-matter density profile.  

Instead, \mhalo can be inferred from an AM model and the more easily measured total stellar mass. While \mstarmhalo relationships give the median stellar mass for a fixed halo mass, by modeling the conditional probability function P(\mstar|\mhalo), taking the inverse of the function does not yield median halo masses when there is scatter in the relationship.  In other words, P(\mhalo|\mstar) $\neq$ P(\mstar |\mhalo), but P(\mhalo |\mstar) $\propto$ P(\mstar |\mhalo) P(\mhalo). A greater preponderance of lower mass halos which upscatter in stellar mass than higher mass halos which downscatter means the true median \mhalo is less than that suggested by the AM relationship. 

We compute the unnormalized P(\mhalo | \mstar) in logarithmic intervals by multiplying the host halo mass function, $\rm{d}n/\rm{d}\log{M}$, by the fraction of halos that are luminous (to account for reionization), and again by the likelihood of it hosting \mstar given an AM model and Gaussian distributed scatter. For the mass functions, we use the form from \cite{Sheth02} and use a transfer function from \cite{Eisenstein98}. Given a fixed stellar mass, we assign relative probabilities to halo masses on a log-scale, then normalize the distribution and find the $50^{\rm{th}}$ percentile. This represents the median expected \mhalo for a fixed \mstar. Accounting for reionization changes the median expected \mhalo by $< 1\%$. The only exception is in the GK16 model when $M_* < 6 \times 10^6$\msun, in which case reionization causes the predicted \mhalo to increase by more than $1\%$ relative to the case of no reionization.

For the GK16 model, this results in halo masses that are $16-34\%$ smaller than the value inferred by inverting the \mstarmhalo relationship. For the GK14 model, the authors estimate a $0.2$ dex lognormal scatter, which results in a $6-10\%$ reduction. The Moster and Brook models do not mention any scatter, but for consistency we continue to assume a $0.2$ dex scatter, yielding a $6-7\%$ and $5-8\%$ mass reduction respectively. While we assume a scatter to infer halo masses, we do not implement a scatter in assigning stellar mass to satellites in any model but GK16 since it is the only one which explicitly takes scatter into account when finding the best fit \mstarmhalo function.

\section{Results}
\label{sec:results}
In the following subsections we compute the mean expected number of luminous satellite galaxies, \nbarlum, above a given stellar mass threshold, $M_*^{\rm{thresh}}$, and the probability of at least one satellite existing above $M_*^{\rm{thresh}}$ as a function the host galaxy's stellar mass.  For the rest of the paper, we consider galaxy stellar masses above \mstar $> 10^3$\msun, since smaller ``hyperfaint'' galaxies have so few stars that they will be difficult to detect above the background. All values are found using the methodology presented in Section~\ref{sec:methods}, generating $30000$ random realizations of satellite populations per host. Due to uncertainty in mass to light ratios, we strictly report on stellar mass, not stellar luminosity.

\subsection{Validation of our models with the Milky Way satellite system}
\label{sec:milky_way}

\begin{figure}
\includegraphics[width=0.48\textwidth]{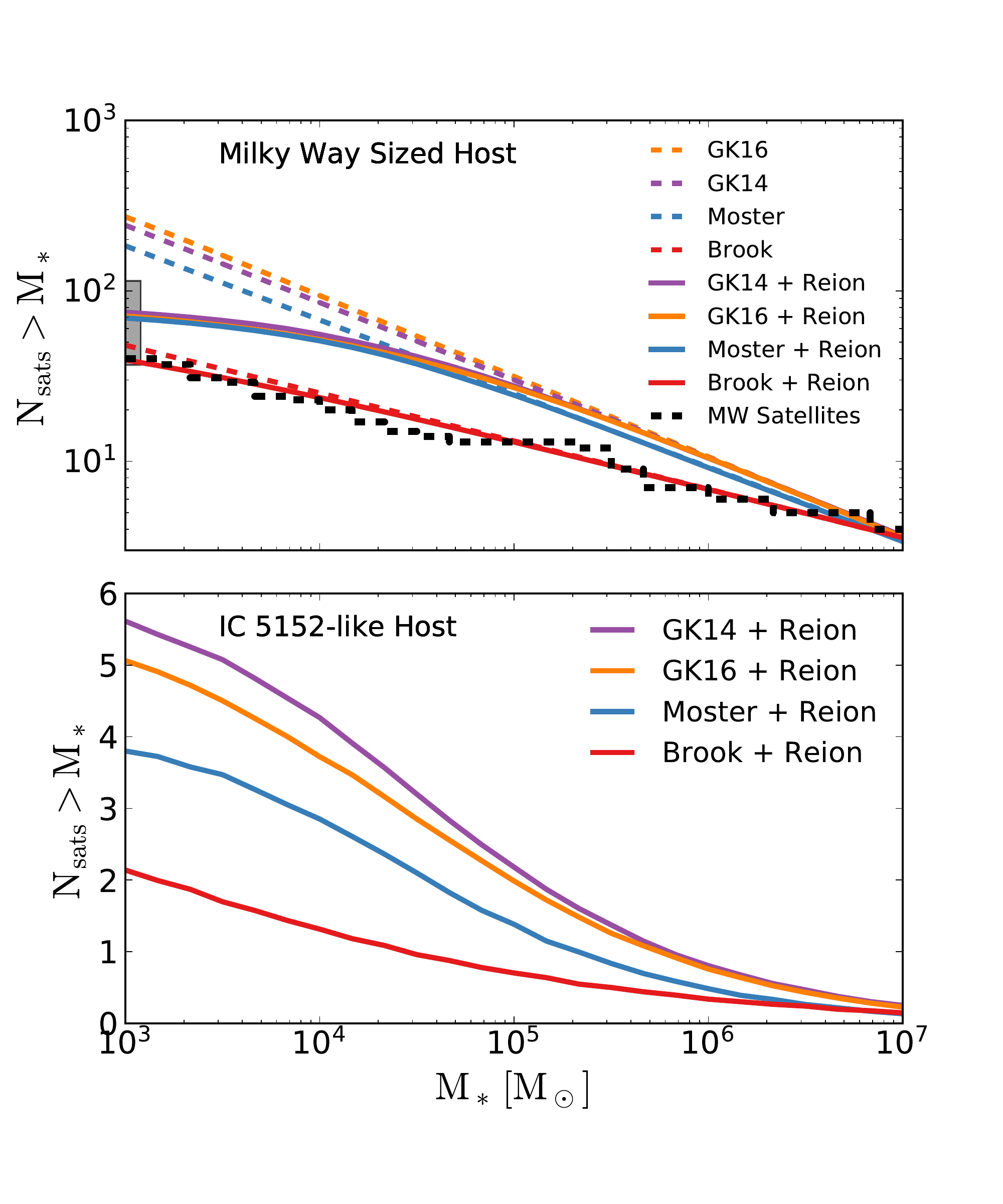}
\caption[Number of satellites around a MW-sized and IC 5152-sized host as a function of satellite stellar mass]{Upper Panel: Mean number of satellites around a MW-sized host with total mass $1.4 \times 10^{12}$\msun as a function of satellite stellar mass. While the Moster, GK14, and GK16 models predict similar numbers of satellites after accounting for reionization, the Brook model predicts significantly fewer. Due to its steep \mstarmhalo relationship, the Brook model is also minimally affected by reionization whereas the other models are reduced by more than $60\%$ on the low mass end. The cumulative stellar mass function of currently known MW satellites is also plotted for comparison. While the Brook model fits the MW satellites the best for higher mass satellites, many more lower mass satellites are expected to be found, which will make it under predict those satellites. The \cite{Hargis14} prediction of $37-114$ satellites with \lstar$> 10^3$\lsun is indicated with a grey box. Lower Panel: Mean number of satellites around a host of $M_* = 2.7 \times 10^{8}$\msun, approximately that of IC 5152, as a function of satellite stellar mass. While the Moster, GK14, and GK16 models make similar predictions if the host's total mass is fixed, they predict different abundances when stellar mass is fixed.}
\label{fig:ngreater_minlum}
\end{figure}

To verify our model implementations, we predict the number of satellites around a Milky Way-sized galaxy. We fix the host's dark matter mass to $M_{\rm{vir}} = 1.4 \times 10^{12}$\msun and plot the mean number of satellites as a function of $M_*^{\rm{thresh}}$ in the upper panel of Fig.~\ref{fig:ngreater_minlum}. We include predictions of each abundance-matching model with and without reionization to indicate how much reionization suppresses the formation of low mass galaxies. The abundances are all scaled to a radius of $300 \, \mathrm{kpc}$ using Eq.~(\ref{eq:radial_lum}) (presented later in Section~\ref{sec:radial}) since that distance corresponds to published completeness limits for satellite surveys of the Milky Way \citep{Walsh2009,Drlica15}. Also plotted is the stellar mass function of $40$ known satellites of the MW with \mstar$>10^3$\msun, with satellites and stellar masses compiled from \cite{McConnachie12}\footnote{Available online at \url{http://www.astro.uvic.ca/~alan/Nearby_Dwarf_Database.html}}, \cite{Drlica-Wagner16}, \cite{Torrealba16b}, and \cite{Torrealba16a}. As in \cite{Dooley17}, the stellar masses are computed from luminosities assuming a stellar mass to light ratio of one, and luminosities are derived from $V$-band absolute magnitudes, doubling the luminosity to account for $M_{\rm V}$ being the luminosity within the half-light radius.

There are $11$ MW satellites with $M_* > 4.5 \times 10^5$\msun within $300 \, \mathrm{kpc}$ of the Galactic Center \citep{GarrisonKimmel16}; all models are consistent with $11$ such satellites within $1 \sigma$ Poisson errors. Since no model was calibrated exclusively to the MW satellites, and there is uncertainty in the mass of the Milky Way from $\approx 0.5 - 2.5 \times 10^{12}$\msun \citep{Wang15}, an exact agreement with the MW luminosity function is not expected. For instance, while the Brook model prediction is remarkably similar to the Milky Way classical dwarf satellite counts for the fiducial MW halo mass, a lower assumed mass would make the other models more consistent.

On the faint end, the models predict a median of $66$, $71$, $67$, and $37$ satellites with $M_* > 10^3$\msun for the Moster, GK14, GK16, and Brook models respectively. The first three models are in strong agreement with a prediction by \cite{Hargis14}, who estimated $69$ satellites above $10^3$\lsun in the MW out to $300 \, \mathrm{kpc}$ from observations and completeness limits, and a $90\%$ confidence interval of $37-114$ satellites, as indicated with a grey rectangle in Fig.~\ref{fig:ngreater_minlum}. \cite{Drlica15} also make a consistent prediction of $~70$ UFDs in the MW when excluding the sub-substructure of UFDs in the Large and Small Magellanic Clouds. 

If $\geq 70$ UFDs, or more than $80$ satellites with $M_* > 10^3$\msun in total are discovered in the MW in the future, it would strongly disfavor the Brook model as implemented. It is difficult to reconcile the Brook model even if the MW halo is on the massive side. At a MW mass of $2 \times 10^{12}$\msun, the Brook model predicts $53$ satellites, more than $2 \sigma$ less than $80$ where $\sigma$ is taken from the negative binomial distribution in \citep{BoylanKolchin10}. If reionization was shut off entirely, the predicted number of satellites in the Brook model could increase by $20\%$. A more likely maximal increase of $10\%$ would still mean the model predicts the MW has an unusually high number of UFDs for its size. The rarity of LMC and SMC sized systems in a MW-like halo \citep{BoylanKolchin10,Busha11}, though, could be evidence in favor of that argument.

In stark contrast, reionization has a large impact on the Moster, GK14, and GK16 models. In these models it is absolutely necessary to include reionization or else an additional $110$ to $205$ luminous satellites with $M_* > 10^3$\msun would be predicted. Based on the MW comparison, we expect numbers predicted by the Brook model to be on the low end of possibilities for satellites of dwarf galaxies, and values from the other models to be closer to median expectations.

\subsection{How many satellites of dwarfs are there?}
\label{sec:satellite_counts}

In the lower panel of Fig.~\ref{fig:ngreater_minlum} we again show the satellite abundance as a function of minimum $M_*^{\rm{sat}}$, but now fix the host halo stellar mass to $M_* = 2.7 \times 10^{8}$\msun to reflect the largest field dwarf galaxy, IC 5152. Without a direct measurement of total baryonic plus dark matter mass available, we convert from stellar mass to halo mass for each AM model as described in Section~\ref{sec:mstar_to_mhalo}. The radius out to which subhalos are counted is the \rvir associated with the halo's total mass, unlike in the top panel. Whereas the Moster, GK14, and GK16 models make similar predictions when the host's total mass is fixed, as in the upper panel, their predictions diverge when instead the host's stellar mass is fixed, as in the lower panel. This is due to the GK models having a lower \mstarmhalo relationship in the range of field halos, which leads to predicting more massive halos for fixed stellar mass. Furthermore, the GK16 and GK14 models are separated due to the larger scatter in the GK16 model, which leads to the GK16 model preferring smaller host halo masses than does the GK14 model. Explicitly, the total halo mass is $9.0 \times 10^{10}$\msun in the GK14 model, $8.5 \times 10^{10}$\msun in the GK16 model, $5.2 \times 10^{10}$\msun in the Moster model, and $4.6 \times 10^{10}$\msun in the Brook model. The models predict a mean of $\sim 2-6$ satellites with $M_*$ above $10^3$\msun, $\sim 1 - 4$ above $10^4$\msun, and $\sim 1 - 2$ above $10^5$\msun for a galaxy like IC 5152.

\begin{figure}
\includegraphics[width=0.48\textwidth]{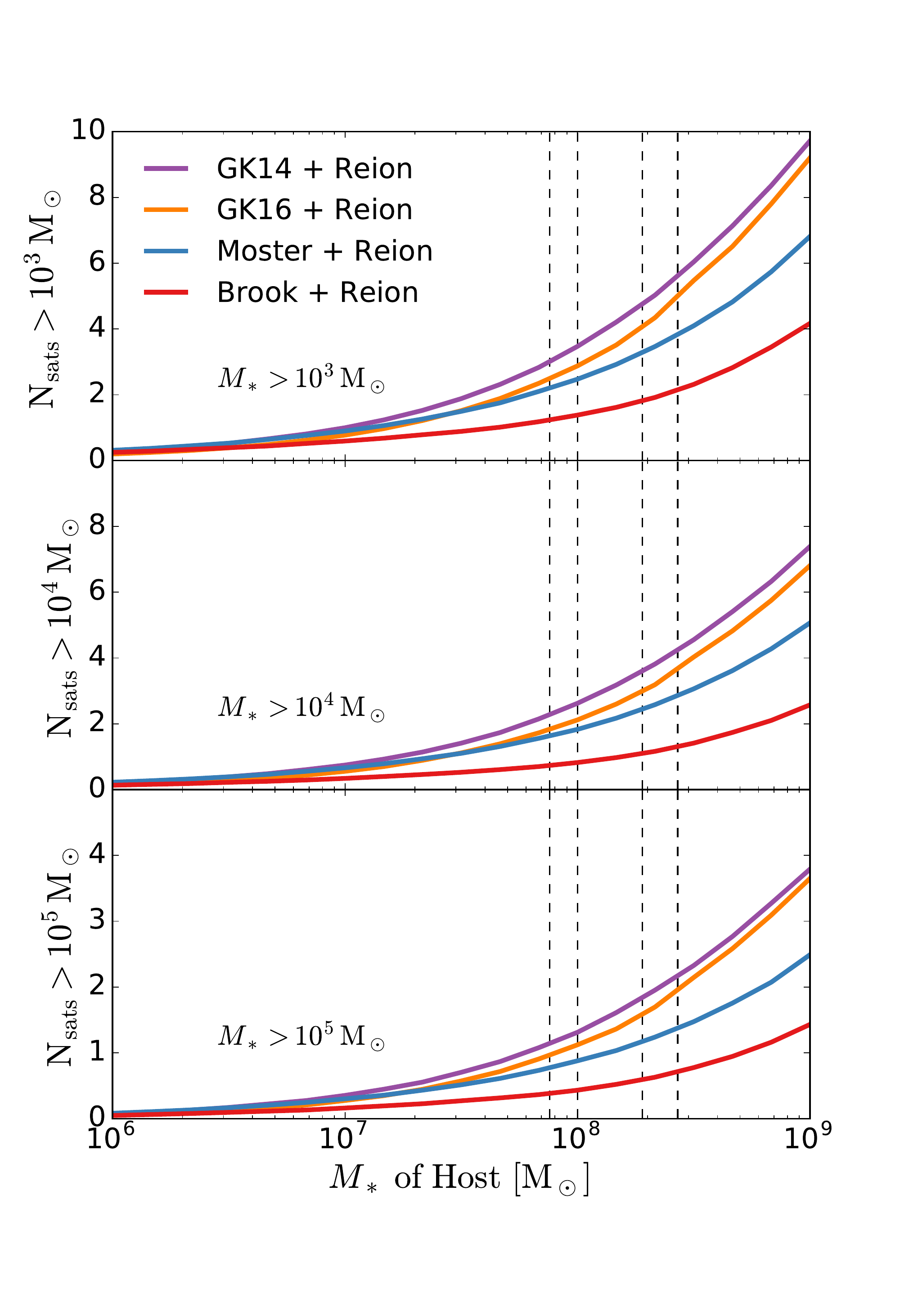}
\caption[Number of satellites as a function of host stellar mass]{Mean number of satellites with stellar mass above $10^3$\msun (upper panel), $10^4$\msun (middle panel), and $10^5$\msun (lower panel) as a function of a host halo's total stellar mass, $M_*$. Each abundance matching model predicts different values, but all agree that satellites should exist around hosts with $M_* \geq 10^8$\msun. Vertical lines correspond to the stellar masses of the five field galaxies listed in Table~\ref{table:results} with the largest stellar masses}.
\label{fig:ngreater}
\end{figure}

In Fig.~\ref{fig:ngreater}, we show the dependence of satellite abundances on the host galaxy's size in terms of its stellar mass. The top panel plots the median number of satellites with $M_*^{\rm{sat}} > 10^3$\msun, the middle panel with $M_*^{\rm{sat}} > 10^4$\msun, and the bottom panel with $M_*^{\rm{sat}} > 10^5$\msun. The dashed vertical lines correspond to the stellar masses of the five largest known field galaxies in the Local Group, IC 5152, IC 4662, IC1613, NGC 6822, and NGC 3109. For IC 5152 for instance, the models predict a mean of $1.3$ satellites with $M_* > 10^4$\msun on the low end in the Brook model, to $4.3$ satellites in the GK14 model.
This figure shows that the number of expected UFDs is a strong function of stellar mass of the host.  Below a host stellar mass of $10^7$\msun, there is a mean of less than one UFD sized satellite according to all models, indicating many hosts will have no satellites above $10^3$\msun. This is broadly consistent with the predictions of \citet{Sales13} and \citet{Wheeler15}, who predicted that the probability of a $10^7$\msun stellar mass host should have a low double-digit percentage probability of hosting an UFD of \mstar$\gtrsim 10^3$\msun. For a host of $10^8$\msun and larger, at least one satellite per host is expected.  

A list of known isolated field galaxies and the mean number satellites within their virial volume is listed in Table~\ref{table:results}. Galaxies and stellar masses were compiled from \cite{McConnachie12}, and supplemented with values from \cite{Karachentsev14}, \cite{McQuinn15}, and \cite{Karachentsev15} for KK 258, Leo P, and KKs 3 respectively. We give an indication of the probabilistic  distribution of satellites by including the $20^{\rm{th}}$ and $80^{\rm{th}}$ percentile of abundance, and the probability that at least one satellite exists, $P(\geq 1)$.

\begin{landscape}
\input Table1
\end{landscape}

\subsection{Likelihood of finding at least one satellite}
\label{sec:probability}
\begin{figure}
\includegraphics[width=0.48\textwidth]{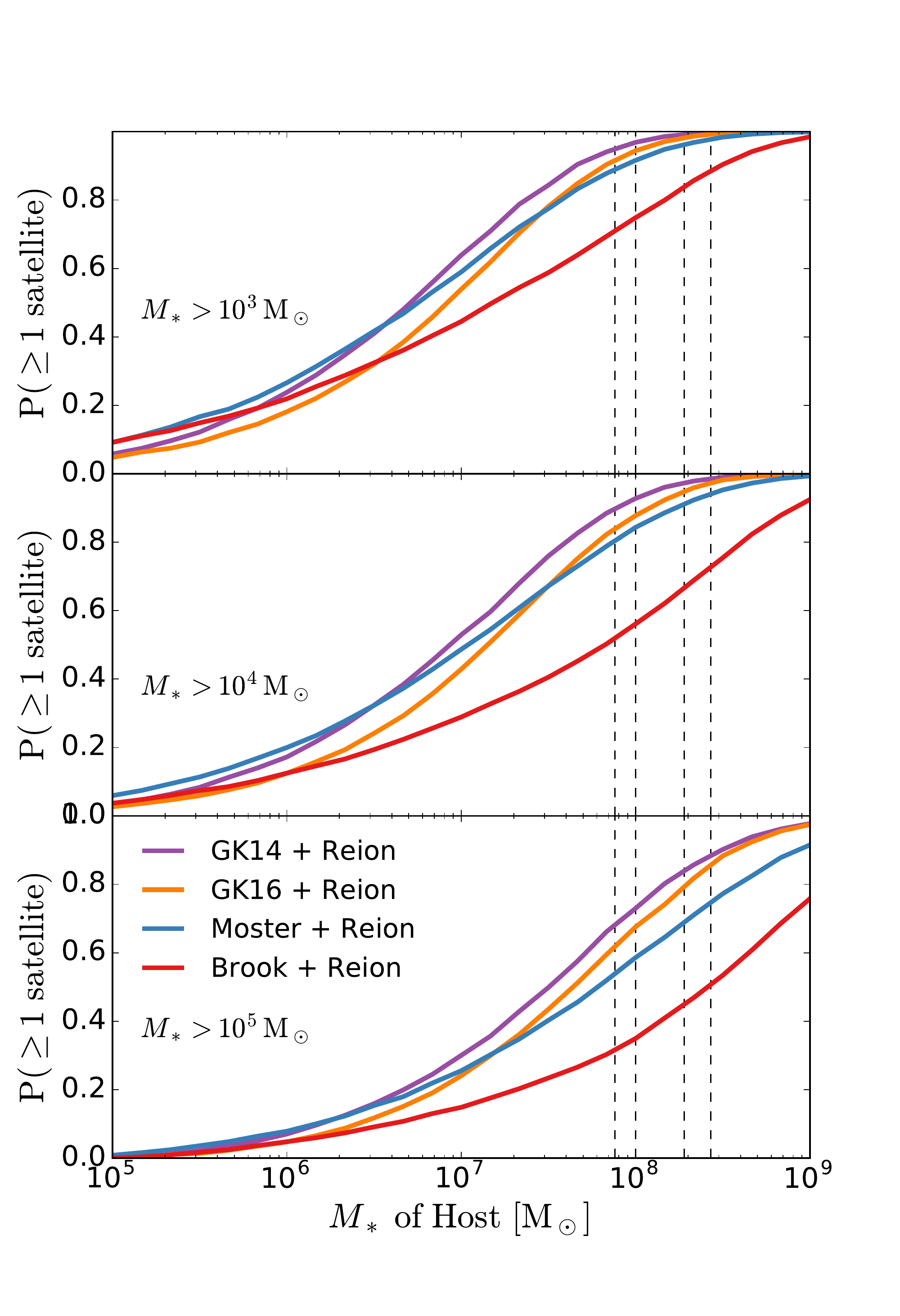}
\caption[Probability of at least one satellite]{Probability that at least one satellite with stellar mass above $10^3$\msun (upper panel), $10^4$\msun (middle pannel), or $10^5$\msun (lower panel) exists around a host with stellar mass $M_*$. The Moster, GK14, and GK16 models all predict a high likelihood of at least one satellite with $M_* > 10^4$\msun existing around each of the five largest field galaxies, whose stellar masses are indicated by dotted lines (and listed in Table~\ref{table:results}). The Brook model predicts lower probabilities, but still $> 50\%$ for each of these galaxies.}
\label{fig:pgt1}
\end{figure}

Another important metric in determining the merit of searching for satellites of field dwarfs galaxies is the probability that at least one satellite exists around a host. In Fig.~\ref{fig:pgt1} we show the probability that at least one satellite with $M_*^{\rm{sat}} > 10^3, 10^4$ and $10^5$\msun exists around a host as a function of \mstar of the host. Dotted vertical lines again show the stellar masses of the five largest known field galaxies in the Local Group. For the largest field galaxy, IC 5152, the probability of a satellite with $M_*^{\rm{sat}} > 10^4$\msun is $> 90\%$ according to the Moster, GK14, and GK16 models. It drops to $73\%$ for the Brook model. For the $5^{th}$ largest field galaxy, the probability remains above $80\%$ in the first three models, and is $51\%$ for the Brook model.

The high likelihoods indicate that a comprehensive search of Local Group dwarf galaxies is likely to yield at least one new satellite discovery, even if only the five most massive field dwarfs are surveyed. For the Moster, GK14, and GK16 models, the probability of one satellite with $M_* > 10^5$\msun is $> 99\%$ if all five of the largest field galaxies are surveyed to their full virial volume. We choose to highlight the five largest because of this fact. In the Brook model, the probability is $92\%$, but goes to $> 99\%$ for $M_* > 10^4$\msun.

\begin{figure}
\includegraphics[width=0.48\textwidth]{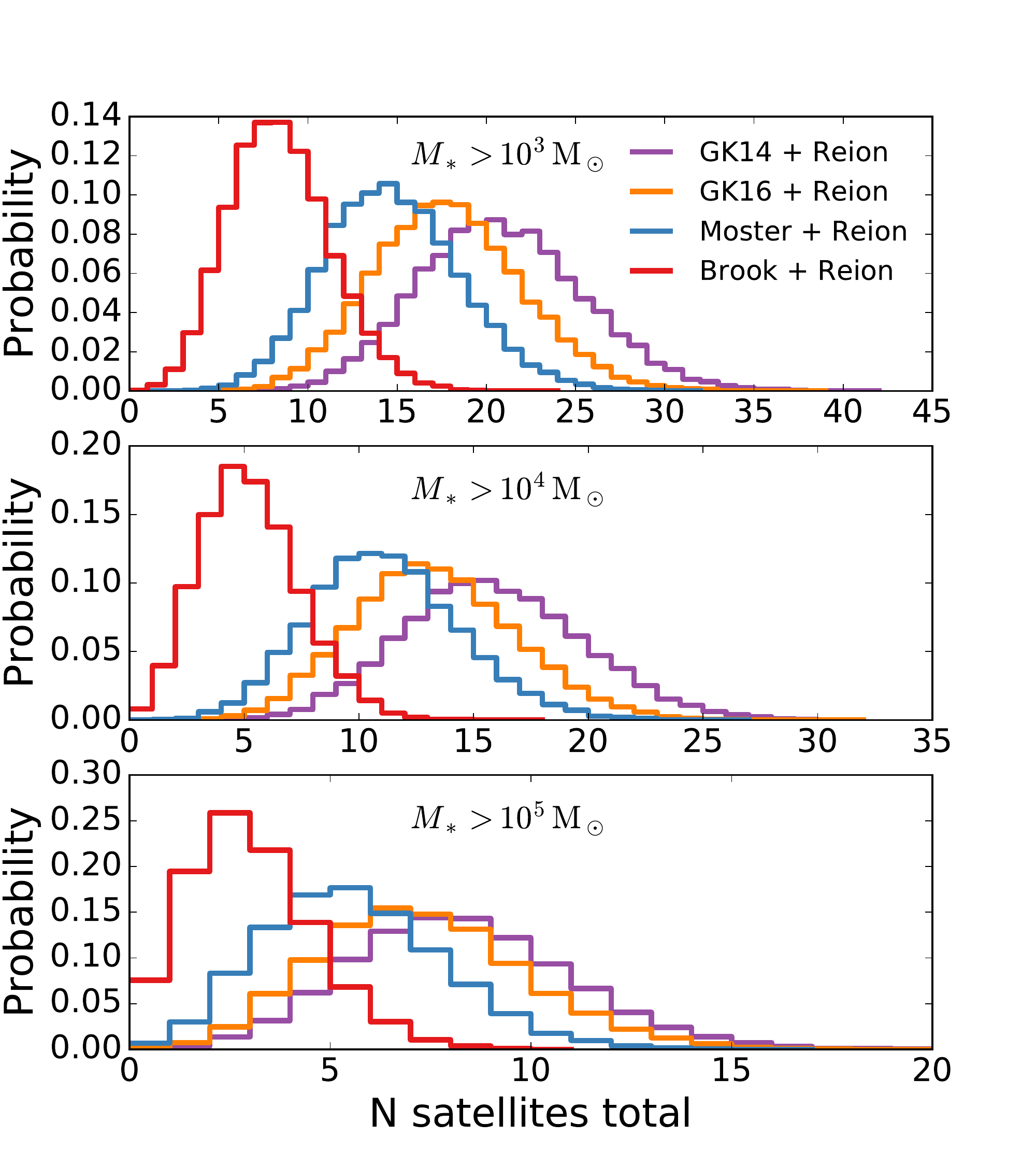}
\caption[Probability distribution of the total number of satellites around select hosts]{Probability distribution for the total number of satellites expected to be found with $M_* > 10^3, 10^4$ and $10^5$\msun around the combined largest five Local Group field galaxies. The GK14 and GK16 models predict over twice the number of satellites as the Brook model, whose predictions are on the lowest end possible to be consistent with MW satellites, and thus are likely a lower limit to the number of satellites of dwarf galaxies. According to the Moster, GK14 and GK16 models, there is a $> 99\%$ that at least one satellite with $M_* > 10^5$\msun exists. These five largest galaxies contain $\sim 1/3$ of the total number of satellites of field dwarfs.}
\label{fig:total_field_sats}
\end{figure}

\begin{figure}
\includegraphics[width=0.48\textwidth]{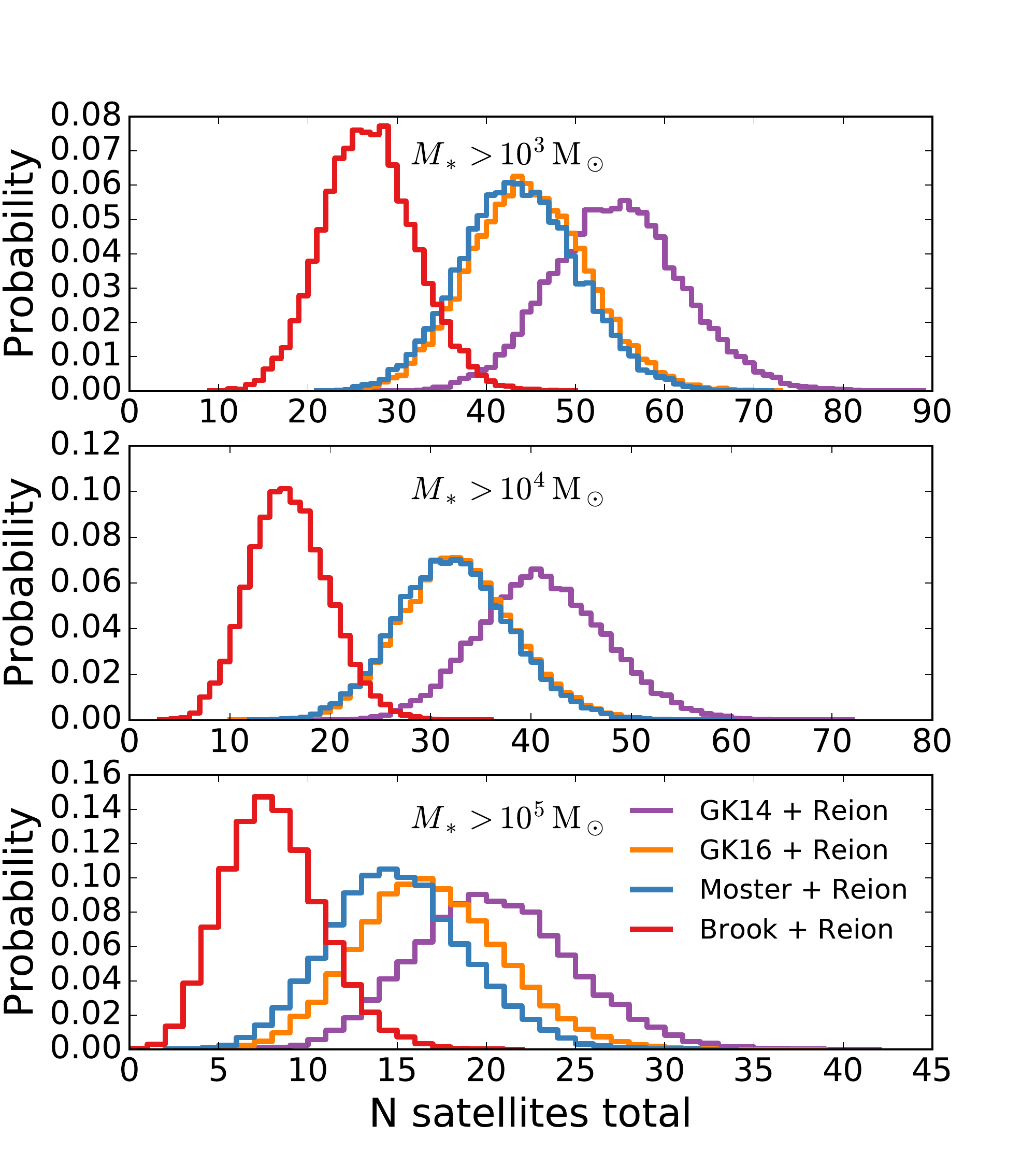}
\caption[Probability distribution of the total number of satellites around all Local Group field galaxies]{Probability distribution for the total number of satellites expected to be found with $M_* > 10^3, 10^4$ and $10^5$\msun around all Local Group field galaxies listed in Table~\ref{table:results}. There is an even greater separation of the Brook and GK14 models than in Fig.~\ref{fig:total_field_sats}. Also in comparison, the GK16 model makes predictions more similar to the Moster model due to a greater scatter in \mstar for the lower mass halos causing the median predicted \mhalo to be less for a given halo stellar mass.}
\label{fig:fulltotal_field_sats}
\end{figure}

In Fig.~\ref{fig:total_field_sats}, we show a probability distribution function for the total number of satellites expected around these five largest field galaxies. The three panels show values for abundances above stellar mass thresholds of $10^3$, $10^4$, and $10^5$\msun. When comparing to Fig.~\ref{fig:fulltotal_field_sats}, which shows the same distribution except for all $38$ field dwarfs listed in Table~\ref{table:results}, we demonstrate that surveying just the five largest galaxies would reveal $\sim 1/3$ of the total population of satellites of field dwarfs. 

The shape of the distributions is driven mostly by Poisson statistics in the number of dark matter subhalos. The randomness of reionization suppressing star formation contributes a smaller additional component. In the case of the Moster model, there is an another perturbation due to differing halo infall times, and in the case of the GK16 model, from scatter in \mstarmhalo. The combined uncertainty arising from counting statistics and abundance matching models results in a broad distribution of possible satellite tallies. The distinct curves from each abundance matching model, in particular comparing the Brook and GK14 models, indicates how observing satellites could provide important insight into ruling out or improving AM models. This is particularly true for satellites above $10^5$\msun since reionization has little effect on the number of satellites above that threshold, and those satellites are easier to discover. Probing down to satellites near $10^3$\msun or $10^4$\msun would help constrain the effects reionization when compared with the number of satellites found above $10^5$\msun, but would be observationally more challenging.



\section{Survey Strategy: Dependence on Field of View}
\label{sec:radial}

Due to the geometry of a line of sight, the values we report in Section~\ref{sec:results} for the number of satellites within a virial volume do not directly translate into expectations for an observed field of view. For instance, the number of dwarfs-of-dwarfs that may exist in the Solo survey depends on their field of view, the distance to target galaxies, the radius of target galaxies, and the radial distribution of satellites within a galaxy. Future campaigns must account for these details when designing a strategy for limited amounts of telescope time. Both targeting larger mass host galaxies and focusing on the innermost region will increase the projected density of satellites. If the observing goal is to discover as many as possible, would it be better to observe the inner region of a lower mass host, or the outer regions of a higher mass host? We address this issue in detail by making predictions specifically for the Solo survey, and discussing a survey strategy. We do not discuss the detectability of dwarf satellites as a function of stellar mass, as this depends on both the intrinsic unknowns of the satellite population (most critically, the distribution function of surface brightness), and survey depth.

We begin by computing a scaling factor to convert the predicted values of the mean number of luminous satellites within a host's virial radius to values expected from surveying circular apertures centered on a host. As found in \cite{Han16} and confirmed in our simulation suite, satellites are distributed approximately spherically symmetrically around a host and the normalized distribution does not vary with host halo size. This allows us to express the normalized cumulative abundance of satellites generically around a host as $K(r)$. We compute $K(r)$ from the subhalos which are deemed luminous in our reionization model across all $33$ \textit{Caterpillar} simulations at $z=0$. Selecting only luminous satellites is crucial, because they are more centrally concentrated than the full sample of dark matter subhalos, as previously found and discussed in \cite{Gao04}, \cite{Starkenburg13}, \cite{Barber14}, and \cite{Sawala16}.

\begin{figure}
\includegraphics[width=0.48\textwidth]{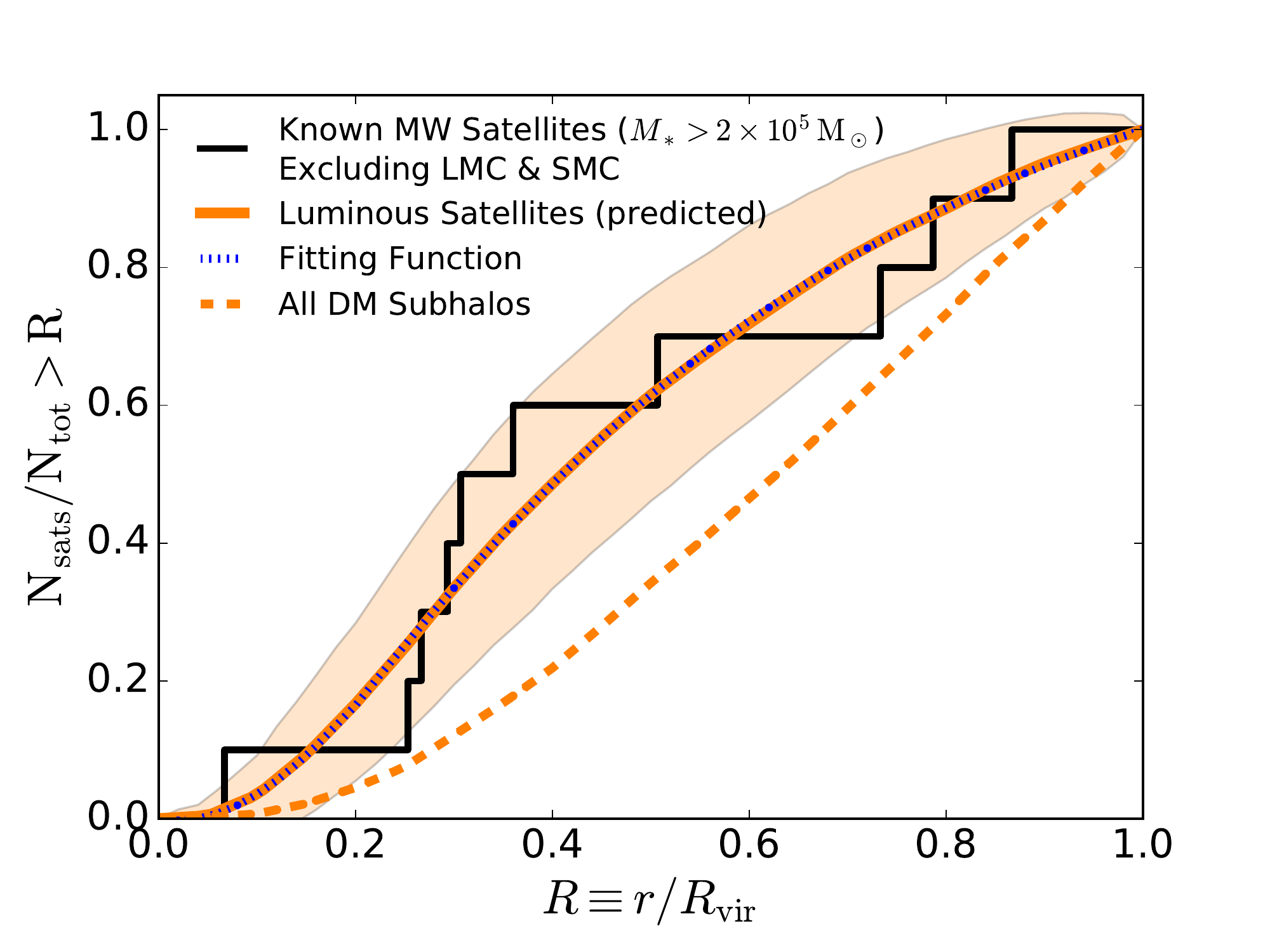} 
\caption[Radial distribution of luminous satellites]{Normalized radial distribution of satellites scaled to a host with \rvir$=300$~kpc. The radial distribution of satellites which survive reionization in our simulations (and are thus luminous) agree well with the radial distribution of known MW satellites, whereas the radial distribution of all dark matter subhalos does not. The sample of known MW satellites includes all satellites with stellar mass above the completeness limit of $M_* = 2 \times 10^5$\msun, except for the LMC and SMC which are known to be spatially correlated and very rare. One sigma variation about the prediction for luminous satellites is shown with a shaded band. It is important to take reionization into account when predicting the radial distribution of satellites, and to not assume it follows the distribution of all dark matter subhalos. A fitting function for the predicted radial distribution is given by equation~(\ref{eq:radial_lum}) and plotted with a dotted line.}
\label{fig:radial_distr_plot}
\end{figure}

The distribution of all dark matter subhalos and the subset of luminous satellites is plotted in Fig.~\ref{fig:radial_distr_plot}. For comparison, we also plot the cumulative distribution of known MW satellites with stellar mass above the observed completeness limit of $M_* = 2 \times 10^5$\msun that are within $300$ kpc of the Galactic Center. We exclude the LMC and SMC since they occur rarely in MW-sized galaxies \citep{BoylanKolchin10,Busha11} and they are spatially correlated. The positions and stellar masses of satellites were taken from \cite{McConnachie12}. Our predicted radial distribution of satellite galaxies fits remarkably well to the MW satellites. In contrast, the distribution inferred from all dark matter subhalos does not, demonstrating the importance of subhalo selection effects due to reionization.

For ease of use, we find a very tight match to the data with a piecewise analytic function. It takes the form
\begin{equation}
\label{eq:radial_lum}
K(R) = 
\begin{cases} 
  k_1 R + k_2 R^2 + k_3 R^3  & R < 0.2 \\
  k_4 \arctan(\frac{R}{k_5} - k_6) & 0.2 \leq R \leq 1.5 \\
\end{cases}
\end{equation}
with best fit values of the constants $k_1 = -0.2615$, $k_2 = 6.888$, $k_3 = -7.035$, $k_4 = 0.9667$, $k_5 = 0.5298$, and $k_6 = 0.2055$. We find that luminous satellites with a lower peak mass are slightly more centrally concentrated than more massive luminous satellites since they are subject to greater selection effects from reionization. However, the shift in $K(R)$ is $\lesssim 10\%$ which is small compared to Poisson noise and much smaller than the difference between all subhalos and only luminous subhalos. Since galaxies are self-similar, and the radial distribution varies only weakly with satellite mass range, it is possible to multiply $K(R)$ by the expected number of satellites in any mass interval within the virial radius of a host (taken from Fig.~\ref{fig:ngreater} for instance) to yield the number of satellites within a radius $r / R_{\rm{vir}}$.

By integrating the density function $\frac{1}{4 \pi r^2} \frac{\rm{d}K}{\rm{d}r}$ over a cylinder of radius $R \equiv r/$\rvir and half depth $Z \equiv z/$\rvir where $r$ and $z$ are cylindrical coordinates centered on the host galaxy, one gets the number of satellites expected in a line of sight relative to the number within \rvir. We call this quantity $K_{\rm{los}}(R)$ We numerically integrate the function and show the result in Fig.~\ref{fig:K(R,Z)}, fixing $Z$ to a value of $1$, approximately where halos are likely to be bound to the host halo, and $1.5$. $Z = 1.5$ encompasses the ``splashback'' radius for slowly accreting halos, defined as the distance to first apocenter of orbiting bound satellites \citep{More15}. It also represents a distance beyond which the density of additional satellites diminishes rapidly towards zero. Since galaxies at all values of $z$ are in the line of sight, $Z=1.5$ is a more accurate reflection of what satellites can be observed.

\begin{figure}
\includegraphics[width=0.48\textwidth]{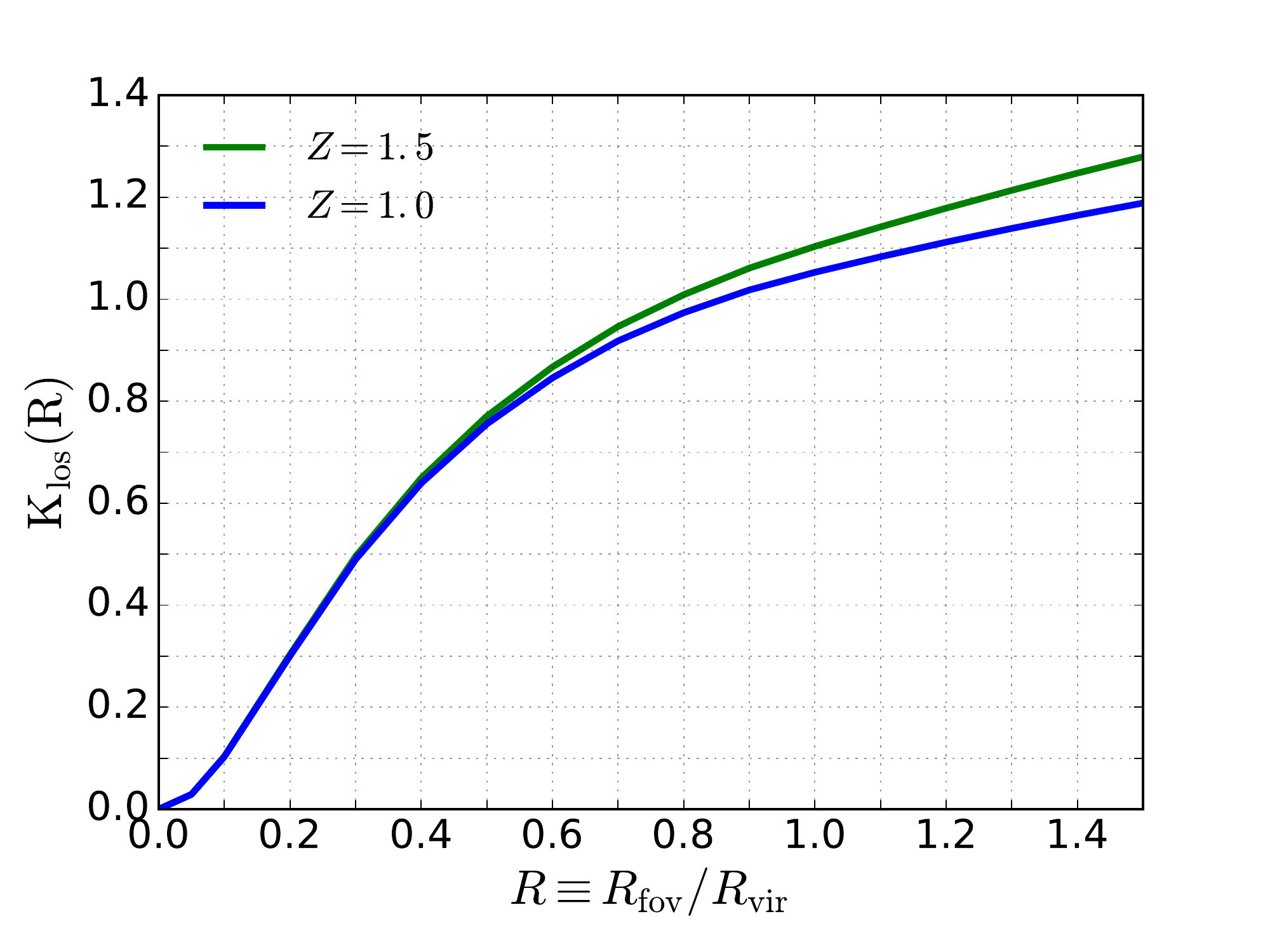} 
\caption[Scaling factor for the number of satellites in a line of sight]{Multiplicative value used to scale the expected number of satellites within a halo's virial radius to the expected number of satellites within an observed column whose field of view has the radius $R_{\rm{fov}}$ on the sky at the distance of the target host galaxy. The column is centered on the galactic center and has a depth of $2 \times Z \equiv 2 \times z/$\rvir. A value of $Z=1$ means the center of the column extends from the near edge of the virial radius to the far edge. For instance, observing a field of view that has $R = 1/2, Z=1$ corresponds to a multiplicative factor of $0.76$.}
\label{fig:K(R,Z)}
\end{figure}

There are several important results from this exercise. First, if one were to search out to the projected virial radius of the host, one would find $\sim 10\%$ more dwarfs than one would expect based on the number of dwarfs strictly within the spherical virial volume. This comes from dwarfs inhabiting a region outside the virial radius that can be imagined by circumscribing a sphere by a cylinder, and then extending the depth (height) of the cylinder to the splashback radius. Second, the number of expected dwarfs rises sublinearly with aperture radius for $R > 0 .5$, regardless of the depth of the line-of-sight, $Z$, under consideration. This is in contrast to the survey area, which grows as the square of the aperture radius. Therefore, in the absence of increased backgrounds like an extended stellar halo, which is unlikely to be significant for dwarfs \citep{Pillepich14}, pointings centered on the host galaxy will have higher satellite yields than individual pointings in the outskirts of the host halos.

In light of this, the most efficient strategy for finding dwarf-of-dwarf satellites is to first target the innermost region of the most massive dwarf hosts. Additional observations should map out the whole virial volume of the biggest dwarfs before targeting down the field dwarf stellar mass function, 
unless environment is a strong motivating factor in the dwarf-of-dwarf search. 

What is the transition between a ``big" and a ``small" dwarf host?  The optimal transition point from searching the innermost region of smaller galaxies to the whole volume of larger galaxies depends on which galaxies are targeted and the telescope's field of view. It can be calculated from Fig.~\ref{fig:K(R,Z)} and Table~\ref{table:results2}. Similar to $K(R)$, $K_{\rm{los}}(R,Z)$ can be multiplied by the number of satellite galaxies within the virial radius of any host, for any satellite mass interval, to yield the number of satellites within a specified line of sight. Table~\ref{table:results2} lists the distances to all isolated field galaxies, repeats the mean number of galaxies within the virial radius, $\bar{N}_{\rm{lum}}$, for convenience, and lists the virial radius of each galaxy inferred from the AM models. \rvir refers to the \cite{Bryan98} definition, consistent with the radius used to determine all SHMFs. We infer \mhalo from the AM model and host halo's stellar mass, then convert \mhalo to \mvir assuming an NFW halo density profile, and finally find \rvir from \mvir.

As an example of computing the number of galaxies in a field of view, IC 5152 is located at a distance of 1.7 Mpc \citep{Zijlstra99}. DECam has a $2.2^{\circ}$ field of view, resulting in a $33 \, \rm{kpc}$ observed radius. At a Moster-model-estimated total halo mass of $5.2 \times 10^{10}$\msun, IC 5152 has a virial radius of $101 \, \rm{kpc}$. Using Fig.~\ref{fig:K(R,Z)} with $Z=1.5$, the number of satellites reported for the virial volume is multiplied by $0.54$ to yield $1.5$ expected satellites in the line of sight. An equivalent calculation for the GK16 model produces $1.8$ expected satellites, $2.0$ for the GK14 model, and $0.7$ for the Brook model. If we were to survey the whole virial volume of IC 5152, we would require an additional 8 distinct pointings, and would expect to find only $1-2$ more satellites. This example also demonstrates that part of the disparity between the Moster, GK14, and GK16 models is generated from predicting different virial volumes for a host with fixed stellar mass. When normalizing to more equivalent volumes, as exemplified here, their predictions start converging.

We apply the same calculations to all of the galaxies in Table~\ref{table:results2} using an approximate field of view from the Solo dwarf survey. Solo has a $1^\circ \times 1^\circ$ camera \citep{Higgs16}, which we approximate as a circular aperture with radius $0.56^\circ$. We present all of the values expected for Solo in the $\bar{N}_{\rm{fov}}$ column. Our table contains all but two of the galaxies they target, Perseus and HIZSS3A(B), for which we could not obtain stellar masses. Due to a small field of view, nearly all galaxies have a mean expected number of satellites of fewer than $1$. However, enough galaxies were surveyed that in aggregate multiple satellites with $M_* > 10^4$\msun could be in their observations. 

One may take even a broader view on host properties than we have so far, asking the question of how one should allocate finite telescope time in order to identify small dwarf galaxies.  The answer to this question depends on one's science goals.  If the goal is simply to discover a lot of low-luminosity, low-surface-brightness objects with a small number of pointing, targeting the central regions of galaxy clusters is the way to go if the required depth can be achieved.  Indeed, this is the approach taken by a number of groups, although the number of \mstar$\lesssim$ a few $\times10^5$\msun dwarfs discovered so far is a literal handful, in part on account of the large distances to even nearby clusters like Virgo \citep{Jang14,Grossauer15,Munoz15,Lee17}.  If the goal is, as we argue, to discover small dwarfs relatively untouched by environmental processes, it is more efficient to target the inner regions of low-mass hosts.  However, because of our incomplete understanding of patchy reionization and environmental processing as a function of host property, it would also be beneficial to identify a field dwarf sample, although such searches require more survey area for a fixed number of desired dwarfs.  A truly comprehensive view of the abundance and star-formation histories of dwarf galaxies requires searches across a range of environments.


\section{Systematic Uncertainties}
\label{sec:sensitivity}
Here we determine what the largest sources of uncertainty are for our luminous dwarf-of-dwarf number count predictions. We show some of them in Fig.~\ref{fig:total_field_sats}: differences between AM models, counting statistics, and halo-to-halo variations. However, there are other sources of uncertainty that we quantify in this section. We introduce \nbarlumfive which is the mean total number of satellites with $M_* > 10^4$\msun that exist around the five largest field galaxies according to the Moster model. Using \nbarlumfive as a baseline, the GK14 model predicts $45\%$ more satellites than the Moster model, and the Brook model predicts $55\%$ less. Statistical fluctuations, driven mostly by Poisson noise, contribute a standard deviation of $\pm 30\%$. Not accounted for are systematic uncertainties in the stellar mass of the host, the total mass of the host, the SHMF, the infall time distribution for the Moster model, the magnitude of scatter in \mstarmhalo for the GK16 model, and the reionization model used. Each of these uncertainties must be estimated before a robust prediction can be made regarding whether or not satellites of dwarfs could be discovered, and how many. We therefore estimate typical uncertainties in each input variable and the resultant effect of uncertainty on \nbarlumfive. Finally, we compare which uncertainties contribute most.

Table~\ref{table:systematic} lists each input variable and variation along with the corresponding $\%$ change in \nbarlumfive. The first value reports the $\%$ change in the number of satellites with $M_* > 10^4$\msun, and the second value in parenthesis is for $M_* > 10^3$\msun. When considering scatter in \mstarmhalo, \nbarlumfive refers to the GK16 model, otherwise all instances refer to the Moster model.

Uncertainty in the total halo mass, $M_{\rm{halo}}^{\rm{host}}$, is computed from the likelihood distribution for \mhalo given \mstar as described in Section~\ref{sec:mstar_to_mhalo}. For the Moster model, one $\sigma$ uncertainty in the largest field dwarf is $+ 22\%$ and $-18\%$. Uncertainty in the GK14 and Brook models is similar, but in the GK16 model which has higher scatter in \mstarmhalo, one $\sigma$ is $+28\%$ and $-35\%$. Consistent with $\bar{N}$ being directly proportional to $M_{\rm{halo}}^{\rm{host}}$ to first order, as written in equation~(\ref{eq:N}), a $22\%$ change in halo mass results in a $23\%$ change in \nbarlumfive.

For stellar mass, we choose an uncertainty of $25\%$, representative of the uncertainty ranges presented in \cite{Roediger15}. Due to the steep dependence of $M_*$ on \mhalo in the AM models, a change in $M_*$ results in a smaller per cent change in the inferred \mhalo. As a result, a $25\%$ change in $M_*$ yields just a $10-12\%$ change in \nbarlumfive, whereas the same change in $M_{\rm{halo}}^{\rm{host}}$ yields a $26\%$ change.


To estimate uncertainty in the infall times of satellites and the SHMF of field halos, we use a jackknife method with one simulation removed at a time. We find the mean infall time of satellites counting back from $z=0$ to be $t_{\rm{infall}}= 7.45 \pm 0.23$ Gyr for the field halos. Shifting the entire infall time distribution by $0.23$ Gyr results in a negligible $<1 \%$ effect on \nbarlumfive. For the SHMF, $\alpha = 1.81\pm .065$. We therefore adjust $\alpha$ to $1.88$ and $1.75$ and find the best fit value of $K_0$ for each slope. A steeper slope reduces $M_*$ at the low \mhalo end, causing fewer expected luminous satellites. A shallower slope does the opposite, with a total uncertainty of $\sim \pm 11\%$.

Next, we change the level of scatter in the GK16 model. A value of $\gamma = 0$ produces a constant lognormal scatter about \mstarmhalo of $\sigma_{\rm{scat}} = 0.2$ dex. Increasing $\gamma$ to $-0.5$ increases the level of scatter relative to our baseline of $\gamma = -0.2$. Three different effects influence the overall outcome of modifying scatter. First, increasing scatter leads to smaller inferred halo mass for a fixed stellar mass, and thus fewer satellites. Second, in GK16's model, increasing scatter requires a steeper \mstarmhalo slope, which reduces $M_*$ for a fixed \mhalo, which subsequently reduces our predicted \nbarlum. Third, increasing scatter causes more of the more numerous lower mass halos to upscatter above a detection threshold than higher mass halos to downscatter, which increases \nbarlum. However, this third effect diminishes on the mass scale where reionization suppresses low mass galaxies. The interplay of all three effects is complex and
\begin{landscape}
\input Table2
\end{landscape}
\noindent model dependent, but ultimately results in less change to \nbarlumfive than exists between abundance matching models.

\begin{figure}
\includegraphics[width=0.48\textwidth]{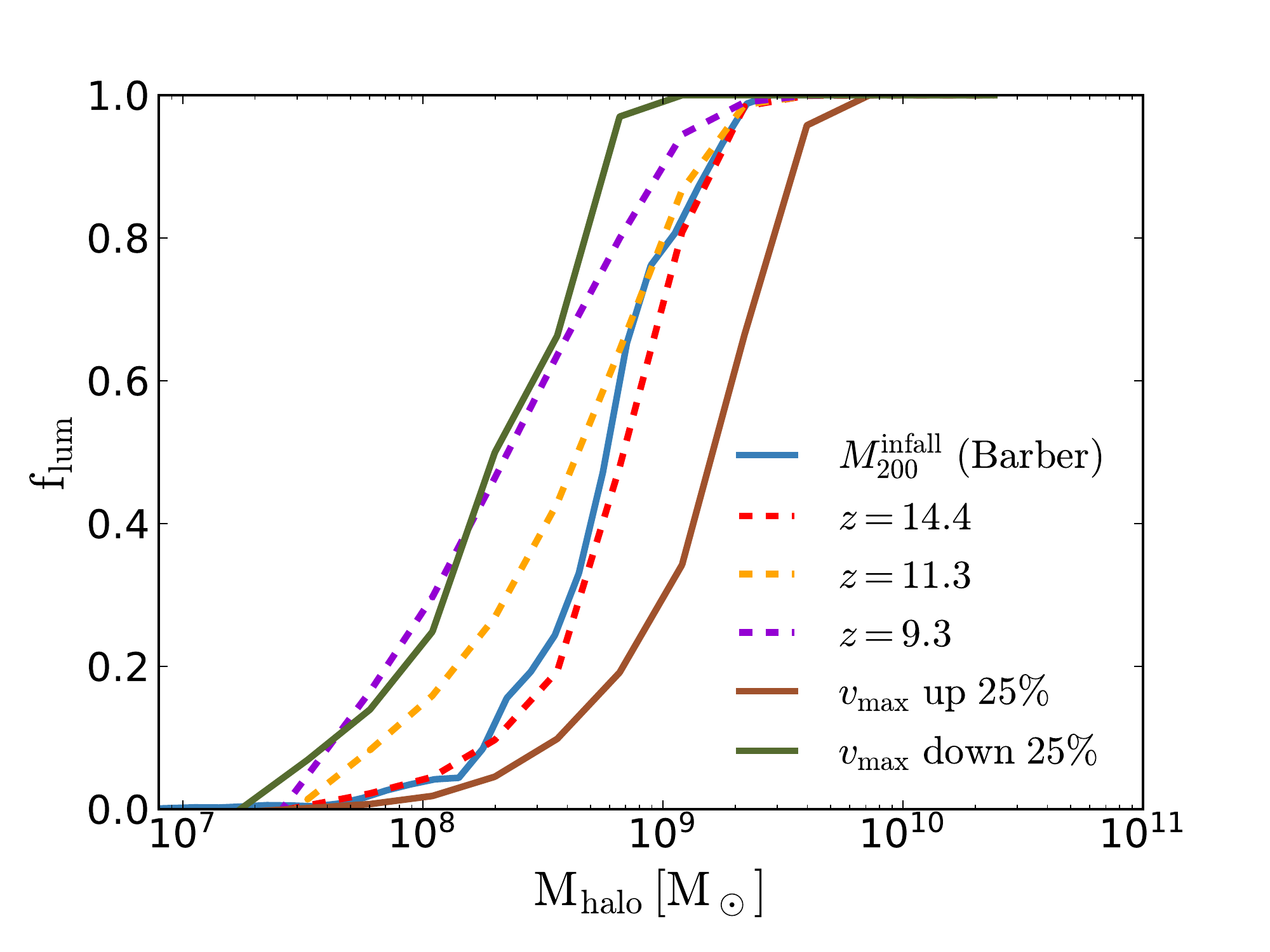} 
\caption[Galaxy luminosity fraction for range of reionization models]{Fraction of dark matter halos that host luminous galaxies at $z = 0$ as a function of their mass \mtwo at infall for various perturbations of the reionization model. The effects of changing the redshift of reionization are explored, as well as changing the maximum circular velocity needed for halos to form stars before and after reionization. Delaying the onset of reionization (lower $z$) and reducing the \vmax needed for a halo to form stars both result in a higher fraction of luminous galaxies. Table~\ref{table:systematic} quantifies these effects relative to the baseline model from \cite{Barber14} (blue, solid line).}
\label{fig:lum_frac_multi}
\end{figure}

Lastly, we consider modifications to reionization. Using the model presented in Section~\ref{sec:reionization}, we are able to adjust the redshift of reionization to $z=14.3$, $11.3$, and $9.3$ and catalog how \nbarlumfive responds in Table~\ref{table:systematic}. We additionally adjust both \vmax thresholds (\vmaxpre and \vmaxfilt) up by $25\%$, both of which create more dark halos, and down by $25\%$, both of which create more luminous halos. The fraction of luminous halos versus halo mass for each case is shown in Fig.~\ref{fig:lum_frac_multi}.

Modifications to reionization can have enormous implications for the abundance of UFD satellites, but little effect on larger satellites. For instance, shifting reionization to later times, $z=9.3$, increases \nbarlumfive by $65\%$ for $M_*>10^3$\msun. For $M_*>10^4$\msun, it increases it by $20\%$, and for $M_*>10^5$\msun there is only a $1\%$ level effect. The same trends are true for adjusting $v_{\rm{max}}^{\rm{pre}}$ and $v_{\rm{max}}^{\rm{filt}}$. For $M_*>10^3$\msun there can be as high as a $70\%$ increase in satellites, while for $M_*>10^5$\msun the increase is just $13\%$.

For an individual halo, the approximate uncertainty from each input for $M_* > 10^4$\msun is as follows: reionization - $33\%$, total halo mass - $20\%$, stellar mass of host - $11\%$, and SHMF - $11\%$. We consider the scatter in \mstarmhalo as part of the spectrum of AM models. When combined in quadrature, the uncertainty reaches $42\%$, commensurate with the differences between abundance matching models, but less than the Poisson noise of systems with \nbarlum$ \leq 4$. Since uncertainty in total halo mass and stellar mass are not fully correlated from one halo to the next, their contribution can be mitigated by observing a larger sample of galaxies. 

Consequently, the dominant contributors to uncertainty are the abundance matching model, reionization, and Poisson noise. While Poisson noise is uncontrollable, reionization and AM models will improve with future observations and better models. For satellites with $10^3 < M_* < 10^4$\msun, uncertainty in reionization is the single most important model dependent factor. However, for satellites with $M_* > 10^5$\msun, reionization has little influence, and differences between abundance matching models dominate. 

Although uncertainties are large, even the most conservative estimates for the existence of satellites suggest that at least one satellite with $M_* > 10^4$ exists around the largest field dwarf galaxies. For $M_* > 10^5$\msun, the lowest estimate comes from the Brook model with a combined uncertainty from reionization, halo mass, stellar mass, and SHMF of $15\%$ for the number of satellites around the five largest field dwarfs. Here, reionization only makes a $2\%$ contribution on its own. Reducing the mean expected predictions by $15\%$ and including Poisson noise, this lower limit still predicts an $88\%$ chance that at least one satellite with $M_* > 10^5$\msun exists around one of the targets.

\begin{table}
\tablewidth{0.47\textwidth}
\centering
\caption[Systematic errors from input variables in satellite abundance predictions]{Systematic Errors}
\label{table:systematic}
\begin{tabular}{lcc}
\hline
\hline
\textbf{Parameter Change} &  \multicolumn{2}{c}{$\%$ change in \nbarlumfive}  \\
\cline{2-3}\\
 & $M_*>10^4$\msun & $M_*>10^3$\msun \\
\hline
Reionization at $z=14.4$ & $ -5$ & $-12 $ \\
\hline
Reionization at $z=11.3$  & $ 8$ & $26 $ \\
\hline
Reionization at $z=9.3$ & $ 20 $ & $65 $ \\

\hline
$v_{\rm{max}}^{\rm{filt}}$, $v_{\rm{max}}^{\rm{pre}} \uparrow 25\%$ & $ -38 $ & $ -47$ \\
\hline
$v_{\rm{max}}^{\rm{filt}}$, $v_{\rm{max}}^{\rm{pre}} \downarrow 25\% $ & $ 27 $ & $ 70 $ \\
\hline
$M_{\rm{halo}}^{\rm{host}} \uparrow 22\%$ & $ 23 $ & $ 26$ \\
\hline
$M_{\rm{halo}}^{\rm{host}} \downarrow 18\%$ & $ -18 $ & $ 26 $ \\
\hline
$M_*^{\rm{host}} \uparrow 25\%$ & $ 10 $ & $ 10$ \\
\hline
$M_*^{\rm{host}} \downarrow 25\%$ & $ -12 $ & $ -12$ \\
\hline
\minftwo SHMF $\alpha = 1.88$ & $ -12 $ & $-9 $ \\
\hline
\minftwo SHMF $\alpha = 1.75$ & $ 10 $ & $ 8$ \\
\hline
GK16, $\gamma = 0.0$ & $ 24 $ & $ 5 $\\ 
\hline
GK16, $\gamma = -0.5$ & $ -32 $ & $-24$\\ 
\hline
$t_{\rm{infall}} \uparrow 0.23$ Gyr & $ <1 $ & $ <1 $\\  
\hline
$t_{\rm{infall}} \downarrow 0.23$ Gyr & $ <1 $ & $ <1 $ \\  
\hline
\hline
\end{tabular}
\tablecomments{Systematic uncertainties in model input variables and their effect on the predicted number of luminous satellites. Per cent change is reported on \nbarlumfive, the mean number of satellites with $M_* > 10^4$\msun found around the largest five galaxies for the Moster model. The values in the final column refer to satellites with $M_* > 10^3$\msun. Adjusting scatter, $\gamma$, in the \mstarmhalo relationship applies only to the GK16 model. More negative values of $\gamma$ indicate more scatter. Our baseline reionization occurs at $z=13.3$, our baseline $\alpha$ in the SHMF is $1.81$, and our baseline $\gamma$ is $-0.2$. $v_{\rm{max}}^{\rm{filt}}$ is the maximum circular velocity of halos above which all halos are assumed to have formed stars, regardless of when this $v_{\rm{max}}$ was reached. $v_{\rm{max}}^{\rm{pre}}$ is the maximum circular velocity of halos which, if reached by the redshift of reionization, indicates star formation will proceed.}
\end{table}


\section{Conclusions}
\label{sec:conclusions}
We have made predictions for the number of luminous satellites around galaxies using SHMFs derived from the \textit{Caterpillar} simulation suite, a model for reionization, and four different AM methods. We find the number of satellites as a function of their minimum stellar mass as well as their host galaxy's total stellar mass. We predict a combined $\sim 70$ ultrafaint and classical satellites ($M_*^{\rm{sat}} > 10^3$\msun) around a MW-sized galaxy, consistent with observational expectations. This result is devoid of any missing satellite problem.

For the more massive isolated Local Group dwarf field galaxies, our predictions overwhelmingly indicate that at least one satellite with $M_* > 10^5$\msun exists, and that many exist with $M_* > 10^4$\msun. Specifically, when observing the virial volumes of the five largest field galaxies combined, there is a $> 99\%$ chance of discovering at least one satellite with $M_*>10^5$\msun, when employing the AM models of Moster, GK14, and GK16, each paired with a reionization model. The existence of at least one such satellite is even supported by our most conservative case, i.e. using the Brook model that underpredicts the number of MW satellites by $\approx 2 \sigma$ compared to other AM models. The Brook model still predicts a $92 \pm 4\%$ chance of one satellite with $M_*>10^5$\msun within the combined virial volumes of the five largest field galaxies, IC~5152, IC~4662, IC~1613, NGC~6822 and NGC~3109. If probing down to $M_*^{\rm{sat}}>10^4$\msun, $5-25$ satellites may even exist. 

We therefore conclude that deep, wide-field searches for faint ($10^3 < M_* < 10^6$\msun) satellites around the known, isolated Local Group field dwarf galaxies should result in the discovery of satellites. This bears the opportunity to study satellites that would likely have environments different from any of the known MW or M31 satellites. A smaller mass host implies they would experience reduced ram pressure stripping, reduced tidal forces, and a different local reionization field. An additional implication is that the MW satellites Fornax ($M_* = 2.5\times 10^7$\msun) and Sagittarius \citep[$M_* \sim 10^8$\msun at infall;][]{Niederste-Ostholt2010} more likely than not had one own small satellite galaxy before getting accreted by the MW. Along the same vein, predictions for the numbers of satellites expected around the SMC and LMC will be reported on in a separate paper.

We estimate how many satellites are in an observable line of sight of specific Local Group galaxies as a function of the radius of the field of view. Making specific predictions will help ensure that the most promising targets are observed to avoid spending significant amounts of telescope time on many (potentially unnecessary) pointings to cover the full virial volume of a given field galaxy. We explicitly calculate expected numbers for galaxies included in the Solo dwarf survey \cite{Higgs16}, which has recently observed all isolated dwarf field galaxies within $3$ Mpc of the MW. For instance, a single pointing at IC 4662 with their $1^\circ \times 1^\circ$ camera would cover a volume that encompasses a mean of $\sim 0.9 - 1.2$ satellites with $M_*>10^4$\msun according to the Moster, GK16, and GK14 models. No satellites of dwarfs have been reported yet in the Solo survey, but given our predictions, we expect that some will be found in their fields of view.



In order to maximize chances of discovering as many satellites as possible, we find the best observing strategy to be to search within small radii of the largest galaxies. After all, the density of satellites is much higher towards their galactic centers than the outskirts. In addition, multiple galaxies should be surveyed this way since observing just one galaxy could yield a non-detection of satellites if it were to have an unusually low halo mass for its stellar mass. If results from a complete survey of all field dwarfs would be available, especially if observations would also cover the full virial volumes of many galaxies including the MW and M31, constraints on the epoch of reionization and the various AM models could be derived.

AM models contribute a large $\sim 50\%$ uncertainty relative to the Moster model to the predictions of satellite abundances. Reionization, which preferentially suppresses star formation in low mass galaxies, can contribute up to $70\%$ uncertainty for $M_*^{\rm{sat}} > 10^3$\msun, but only $\sim 10\%$ for $M_*^{\rm{sat}} > 10^5$\msun. Measurements to a completeness limit of $10^5$\msun would primarily constrain the \mstarmhalo relationship for galaxies and thus the accuracy of the abundance matching models. If observations could ever yield a census of satellites down to $10^3$\msun, reionization could also be constrained by placing limits on the ability of reionization to suppress (or not) star formation in these small halos. The combination of AM prescription and reionization model shape the galaxy stellar mass function both globally and in individual systems, and can be inferred from the mass function.  We show a specific example of this in \citet{Dooley17}, in which we find that the stellar mass function of the newly discovered dwarfs in the direction of the LMC suggests both a break in the AM model and late reionization.  Observations of the dwarf satellites of dwarf galaxies are critical to determining if this explanation for the LMC is universal, or if the LMC is an outlier in its satellite stellar mass function.

Finally, we emphasize that discovering no satellites at all is highly unlikely. Nevertheless, a non-detection would imply at least two of the following: a very strong suppression of star formation by reionization, a low \mstarmhalo relationship for low mass galaxies, a high \mstarmhalo relationship for galaxies with masses typical of field dwarfs, and a MW which has an abnormally large number of luminous satellites with $M_* < 10^6$\msun. We therefore conclude that there are almost certainly many small satellites of dwarf galaxies waiting to be discovered and that their discovery will help refine not just AM models and reionization but our understanding of low mass galaxy formation.


\section*{Acknowledgements} 
G.A.D. acknowledges support from an NSF Graduate Research Fellowship under Grant No. 1122374. Support for A.H.G.P. was provided in part by NASA through the HST theory grant HST-AR-13896.005-A from the Space Telescope Science Institute, which is operated by the Association of Universities for Research in Astronomy, Inc., under NASA contract NAS 5-26555. BW acknowledges support by NSF Faculty Early Career Development (CAREER) award AST-1151462. We thank Shea Garrison-Kimmel for private correspondence regarding abundance matching implementations, and Chris Barber for providing data we used for our default reionization model.

\label{lastpage}
\end{document}

%% file: Table1.tex
\begin{table}
\tablewidth{0.97\textwidth}
\centering
\caption{Mean number of satellites with $M_* > 10^4$\msun around Local Group dwarf field galaxies}
\label{table:results}
\begin{tabular}{lcccccccccccccccc} 
\hline
\hline
\textbf{Name} &  \boldmath$M_* $ & \multicolumn{3}{c}{\textbf{Moster}} && \multicolumn{3}{c}{\textbf{GK14}} && \multicolumn{3}{c}{\textbf{GK16}} && \multicolumn{3}{c}{\textbf{Brook}} \\
 \cline{3-5}\cline{7-9}\cline{11-13} \cline{15-17} \\ 
  & $[10^6 \, \mathrm{M_\odot}]$  & $\bar{N}_{\rm{lum}}$ & 20/80\% &  $P( \geq 1)$ &   & $\bar{N}_{\rm{lum}}$ & 20/80\% & $P( \geq 1)$ & & $\bar{N}_{\rm{lum}}$  & 20/80\% & $P( \geq 1)$ &  & $\bar{N}_{\rm{lum}}$ & 20/80\% & $P( \geq 1)$  \\
\hline
Leo T & 0.14      & $0.08$ & $0 / 0$ & 0.07 &    & $0.05$ & $0/ 0$ & 0.05 &    & $0.03$ & $0/0$ & 0.04  &     & $0.05$ & $0/0$ & 0.05    \\
\hline
And XXVIII & 0.21      & $0.10$ & $0 / 0$ & 0.09 &    & $0.07$ & $0/ 0$ & 0.07 &    & $0.05$ & $0/0$ & 0.05  &     & $0.06$ & $0/0$ & 0.06    \\
\hline
KKR 3 & 0.54      & $0.16$ & $0 / 0$ & 0.15 &    & $0.13$ & $0/ 0$ & 0.12 &    & $0.09$ & $0/0$ & 0.09  &     & $0.10$ & $0/0$ & 0.10    \\
\hline
Tucana & 0.56      & $0.17$ & $0 / 0$ & 0.15 &    & $0.13$ & $0/ 0$ & 0.12 &    & $0.09$ & $0/0$ & 0.09  &     & $0.10$ & $0/0$ & 0.09    \\
\hline
Leo P & 0.56      & $0.17$ & $0 / 0$ & 0.15 &    & $0.13$ & $0/ 0$ & 0.12 &    & $0.09$ & $0/0$ & 0.09  &     & $0.10$ & $0/0$ & 0.10    \\
\hline
And XVIII & 0.63      & $0.17$ & $0 / 0$ & 0.16 &    & $0.15$ & $0/ 0$ & 0.13 &    & $0.10$ & $0/0$ & 0.09  &     & $0.11$ & $0/0$ & 0.10    \\
\hline
Phoenix & 0.77      & $0.19$ & $0 / 0$ & 0.18 &    & $0.17$ & $0/ 0$ & 0.15 &    & $0.11$ & $0/0$ & 0.11  &     & $0.12$ & $0/0$ & 0.11    \\
\hline
KKH 86 & 0.82      & $0.19$ & $0 / 0$ & 0.18 &    & $0.17$ & $0/ 0$ & 0.15 &    & $0.12$ & $0/0$ & 0.11  &     & $0.12$ & $0/0$ & 0.11    \\
\hline
Antlia & 1.3      & $0.25$ & $0 / 1$ & 0.22 &    & $0.23$ & $0/ 1$ & 0.20 &    & $0.16$ & $0/0$ & 0.15  &     & $0.15$ & $0/0$ & 0.14    \\
\hline
KKR 25 & 1.4      & $0.26$ & $0 / 1$ & 0.23 &    & $0.24$ & $0/ 1$ & 0.21 &    & $0.16$ & $0/0$ & 0.15  &     & $0.15$ & $0/0$ & 0.14    \\
\hline
Aquarius & 1.6      & $0.28$ & $0 / 1$ & 0.24 &    & $0.26$ & $0/ 1$ & 0.22 &    & $0.18$ & $0/0$ & 0.16  &     & $0.16$ & $0/0$ & 0.15    \\
\hline
DDO 113 & 2.1      & $0.32$ & $0 / 1$ & 0.27 &    & $0.30$ & $0/ 1$ & 0.26 &    & $0.21$ & $0/0$ & 0.19  &     & $0.18$ & $0/0$ & 0.17    \\
\hline
Cetus & 2.6      & $0.35$ & $0 / 1$ & 0.29 &    & $0.35$ & $0/ 1$ & 0.29 &    & $0.24$ & $0/1$ & 0.21  &     & $0.20$ & $0/0$ & 0.18    \\
\hline
ESO 294-G010 & 2.7      & $0.36$ & $0 / 1$ & 0.30 &    & $0.35$ & $0/ 1$ & 0.29 &    & $0.25$ & $0/1$ & 0.22  &     & $0.21$ & $0/0$ & 0.18    \\
\hline
Sagittarius dIrr & 3.5      & $0.41$ & $0 / 1$ & 0.33 &    & $0.41$ & $0/ 1$ & 0.34 &    & $0.30$ & $0/1$ & 0.25  &     & $0.23$ & $0/1$ & 0.20    \\
\hline
ESO 410-G005 & 3.5      & $0.41$ & $0 / 1$ & 0.33 &    & $0.41$ & $0/ 1$ & 0.34 &    & $0.29$ & $0/1$ & 0.25  &     & $0.23$ & $0/1$ & 0.21    \\
\hline
KKH 98 & 4.5      & $0.46$ & $0 / 1$ & 0.37 &    & $0.48$ & $0/ 1$ & 0.38 &    & $0.35$ & $0/1$ & 0.29  &     & $0.25$ & $0/1$ & 0.22    \\
\hline
Leo A & 6      & $0.52$ & $0 / 1$ & 0.41 &    & $0.56$ & $0/ 1$ & 0.43 &    & $0.41$ & $0/1$ & 0.33  &     & $0.28$ & $0/1$ & 0.25    \\
\hline
GR 8 & 6.4      & $0.54$ & $0 / 1$ & 0.41 &    & $0.58$ & $0/ 1$ & 0.44 &    & $0.43$ & $0/1$ & 0.34  &     & $0.29$ & $0/1$ & 0.25    \\
\hline
Pegasus dIrr & 6.6      & $0.54$ & $0 / 1$ & 0.42 &    & $0.59$ & $0/ 1$ & 0.45 &    & $0.43$ & $0/1$ & 0.35  &     & $0.29$ & $0/1$ & 0.25    \\
\hline
UGC 9128 & 7.8      & $0.59$ & $0 / 1$ & 0.45 &    & $0.65$ & $0/ 1$ & 0.47 &    & $0.48$ & $0/1$ & 0.38  &     & $0.31$ & $0/1$ & 0.27    \\
\hline
UGC 4879 & 8.3      & $0.60$ & $0 / 1$ & 0.46 &    & $0.67$ & $0/ 1$ & 0.49 &    & $0.49$ & $0/1$ & 0.39  &     & $0.32$ & $0/1$ & 0.27    \\
\hline
KK 258 & 14      & $0.77$ & $0 / 1$ & 0.53 &    & $0.90$ & $0/ 2$ & 0.59 &    & $0.67$ & $0/1$ & 0.49  &     & $0.39$ & $0/1$ & 0.32    \\
\hline
UGCA 86 & 16      & $0.82$ & $0 / 1$ & 0.56 &    & $0.97$ & $0/ 2$ & 0.62 &    & $0.74$ & $0/1$ & 0.53  &     & $0.41$ & $0/1$ & 0.33    \\
\hline
\end{tabular}
\end{table}
\clearpage
\begin{table}
\tablewidth{0.97\textwidth}
\centering
\contcaption{Mean number of satellites with $M_* > 10^4$\msun around Local Group dwarf field galaxies}
\begin{tabular}{lcccccccccccccccc} 
\hline
\hline
\textbf{Name} &  \boldmath$M_* $ & \multicolumn{3}{c}{\textbf{Moster}} && \multicolumn{3}{c}{\textbf{GK14}} && \multicolumn{3}{c}{\textbf{GK16}} && \multicolumn{3}{c}{\textbf{Brook}} \\
 \cline{3-5}\cline{7-9}\cline{11-13} \cline{15-17} \\ 
  & $[10^6 \, \mathrm{M_\odot}]$  & $\bar{N}_{\rm{lum}}$ & 20/80\% &  $P( \geq 1)$ &   & $\bar{N}_{\rm{lum}}$ & 20/80\% & $P( \geq 1)$ & & $\bar{N}_{\rm{lum}}$  & 20/80\% & $P( \geq 1)$ &  & $\bar{N}_{\rm{lum}}$ & 20/80\% & $P( \geq 1)$  \\
\hline
DDO 99 & 16     & $0.82$ & $0 / 1$ & 0.56 &    & $0.98$ & $0/ 2$ & 0.62 &    & $0.73$ & $0/1$ & 0.52  &     & $0.41$ & $0/1$ & 0.34    \\
\hline
UKS 2323-326 & 17     & $0.84$ & $0 / 2$ & 0.57 &    & $1.01$ & $0/ 2$ & 0.63 &    & $0.76$ & $0/1$ & 0.54  &     & $0.41$ & $0/1$ & 0.34    \\
\hline
UGC 8508 & 19     & $0.88$ & $0 / 2$ & 0.58 &    & $1.07$ & $0/ 2$ & 0.66 &    & $0.82$ & $0/1$ & 0.56  &     & $0.43$ & $0/1$ & 0.35    \\
\hline
KKs 3 & 23     & $0.96$ & $0 / 2$ & 0.62 &    & $1.19$ & $0/ 2$ & 0.69 &    & $0.92$ & $0/2$ & 0.60  &     & $0.47$ & $0/1$ & 0.37    \\
\hline
NGC 4163 & 37     & $1.19$ & $0 / 2$ & 0.70 &    & $1.53$ & $0/ 3$ & 0.78 &    & $1.22$ & $0/2$ & 0.70  &     & $0.55$ & $0/1$ & 0.42    \\
\hline
WLM & 43     & $1.28$ & $0 / 2$ & 0.72 &    & $1.67$ & $1/ 3$ & 0.81 &    & $1.34$ & $0/2$ & 0.74  &     & $0.59$ & $0/1$ & 0.44    \\
\hline
Sextans A & 44     & $1.28$ & $0 / 2$ & 0.72 &    & $1.70$ & $1/ 3$ & 0.82 &    & $1.34$ & $0/2$ & 0.74  &     & $0.60$ & $0/1$ & 0.45    \\
\hline
DDO 125 & 47     & $1.32$ & $0 / 2$ & 0.74 &    & $1.75$ & $1/ 3$ & 0.83 &    & $1.40$ & $0/2$ & 0.76  &     & $0.61$ & $0/1$ & 0.46    \\
\hline
DDO 190 & 51     & $1.37$ & $0 / 2$ & 0.74 &    & $1.84$ & $1/ 3$ & 0.84 &    & $1.47$ & $0/2$ & 0.77  &     & $0.62$ & $0/1$ & 0.47    \\
\hline
Sextans B & 52     & $1.37$ & $0 / 2$ & 0.75 &    & $1.85$ & $1/ 3$ & 0.84 &    & $1.48$ & $0/2$ & 0.78  &     & $0.64$ & $0/1$ & 0.47    \\
\hline
IC 3104 & 62     & $1.50$ & $0 / 2$ & 0.77 &    & $2.04$ & $1/ 3$ & 0.87 &    & $1.64$ & $1/3$ & 0.81  &     & $0.68$ & $0/1$ & 0.49    \\
\hline
NGC 3109 & 76     & $1.63$ & $1 / 3$ & 0.80 &    & $2.26$ & $1/ 3$ & 0.90 &    & $1.81$ & $1/3$ & 0.84  &     & $0.73$ & $0/1$ & 0.52    \\
\hline
NGC 6822 & 100     & $1.84$ & $1 / 3$ & 0.84 &    & $2.62$ & $1/ 4$ & 0.93 &    & $2.12$ & $1/3$ & 0.88  &     & $0.83$ & $0/2$ & 0.56    \\
\hline
IC 1613 & 100     & $1.84$ & $1 / 3$ & 0.84 &    & $2.62$ & $1/ 4$ & 0.93 &    & $2.12$ & $1/3$ & 0.88  &     & $0.83$ & $0/2$ & 0.56    \\
\hline
IC 4662 & 190     & $2.44$ & $1 / 4$ & 0.91 &    & $3.61$ & $2/ 5$ & 0.97 &    & $2.99$ & $1/4$ & 0.95  &     & $1.09$ & $0/2$ & 0.67    \\
\hline
IC 5152 & 270     & $2.83$ & $1 / 4$ & 0.94 &    & $4.25$ & $2/ 6$ & 0.98 &    & $3.73$ & $2/5$ & 0.98  &     & $1.30$ & $0/2$ & 0.73    \\
\hline
\hline
\end{tabular}
\tablecomments{Mean number of satellites with $M^* > 10^4$\msun expected to exist within the virial volume of known Local Group dwarf irregular and dwarf spheroidal galaxies as predicted with various AM models. The $20^{\rm{th}}$ and $80^{\rm{th}}$ percentile of the satellite abundance distributions are included in the second column. Also shown is the probability of finding at least one satellite around each galaxy, $P( \geq 1)$.}
\end{table}

%% file: Table2.tex
\begin{table}
\tablewidth{0.97\textwidth}
\centering
\caption{Mean number of satellites with $M_* > 10^4$\msun within a $0.56^\circ$ radius field of view around Local Group dwarf field galaxies}
\label{table:results2}
\begin{tabular}{lcccccccccccccccc} 
\hline
\hline
\textbf{Name} &  \boldmath$D_{\sun} \, [\mathrm{kpc}]$ & \multicolumn{3}{c}{\textbf{Moster}} && \multicolumn{3}{c}{\textbf{GK14}} && \multicolumn{3}{c}{\textbf{GK16}} && \multicolumn{3}{c}{\textbf{Brook}} \\
 \cline{3-5}\cline{7-9}\cline{11-13} \cline{15-17} \\ 
  &  & $\bar{N}_{\rm{lum}}$ &  $\bar{N}_{\rm{fov}}$ & $R_{\rm{vir}} \, \rm{[kpc]}$  &  & $\bar{N}_{\rm{lum}}$ & $\bar{N}_{\rm{fov}}$ & $R_{\rm{vir}} \, \rm{[kpc]}$ & & $\bar{N}_{\rm{lum}}$  & $\bar{N}_{\rm{fov}}$ & $R_{\rm{vir}} \, \rm{[kpc]}$ &  & $\bar{N}_{\rm{lum}}$ & $\bar{N}_{\rm{fov}}$ & $R_{\rm{vir}} \, \rm{[kpc]}$   \\
\hline
Leo T & 417    & $0.08$ & 0.01 & 34 &    & $0.04$ & 0.01 & 31 &    & $0.03$ & 0.00 & 30 &     & $0.04$ & 0.00 & 41    \\
\hline
And XXVIII & 661    & $0.10$ & 0.02 & 36 &    & $0.06$ & 0.02 & 33 &    & $0.05$ & 0.01 & 32 &     & $0.06$ & 0.01 & 43    \\
\hline
KKR 3 & 2188    & $0.16$ & 0.13 & 41 &    & $0.13$ & 0.10 & 39 &    & $0.09$ & 0.07 & 37 &     & $0.10$ & 0.07 & 48    \\
\hline
Tucana & 887    & $0.18$ & 0.05 & 42 &    & $0.13$ & 0.04 & 40 &    & $0.09$ & 0.03 & 37 &     & $0.10$ & 0.02 & 48    \\
\hline
Leo P & 1620    & $0.17$ & 0.10 & 42 &    & $0.13$ & 0.08 & 40 &    & $0.09$ & 0.06 & 37 &     & $0.10$ & 0.05 & 48    \\
\hline
And XVIII & 1355    & $0.18$ & 0.09 & 42 &    & $0.14$ & 0.08 & 40 &    & $0.10$ & 0.05 & 38 &     & $0.11$ & 0.05 & 49    \\
\hline
Phoenix & 415    & $0.20$ & 0.02 & 43 &    & $0.16$ & 0.01 & 42 &    & $0.11$ & 0.01 & 39 &     & $0.11$ & 0.01 & 50    \\
\hline
KKH 86 & 2582    & $0.20$ & 0.16 & 44 &    & $0.17$ & 0.14 & 42 &    & $0.12$ & 0.11 & 39 &     & $0.12$ & 0.09 & 50    \\
\hline
Antlia & 1349    & $0.25$ & 0.11 & 47 &    & $0.23$ & 0.10 & 46 &    & $0.15$ & 0.07 & 43 &     & $0.15$ & 0.06 & 53    \\
\hline
KKR 25 & 1905    & $0.26$ & 0.16 & 47 &    & $0.24$ & 0.15 & 46 &    & $0.17$ & 0.11 & 43 &     & $0.15$ & 0.08 & 53    \\
\hline
Aquarius & 1072    & $0.28$ & 0.09 & 48 &    & $0.26$ & 0.08 & 47 &    & $0.18$ & 0.06 & 44 &     & $0.16$ & 0.05 & 54    \\
\hline
DDO 113 & 2951    & $0.32$ & 0.26 & 50 &    & $0.31$ & 0.26 & 50 &    & $0.21$ & 0.18 & 46 &     & $0.18$ & 0.14 & 56    \\
\hline
Cetus & 755    & $0.36$ & 0.06 & 52 &    & $0.35$ & 0.06 & 52 &    & $0.24$ & 0.05 & 48 &     & $0.20$ & 0.03 & 57    \\
\hline
ESO 294-G010 & 2032    & $0.36$ & 0.22 & 52 &    & $0.35$ & 0.21 & 52 &    & $0.24$ & 0.16 & 48 &     & $0.21$ & 0.12 & 57    \\
\hline
Sagittarius dIrr & 1067    & $0.39$ & 0.11 & 54 &    & $0.42$ & 0.11 & 54 &    & $0.29$ & 0.09 & 51 &     & $0.23$ & 0.05 & 59    \\
\hline
ESO 410-G005 & 1923    & $0.41$ & 0.23 & 54 &    & $0.41$ & 0.23 & 54 &    & $0.29$ & 0.17 & 51 &     & $0.23$ & 0.12 & 59    \\
\hline
KKH 98 & 2523    & $0.46$ & 0.32 & 56 &    & $0.46$ & 0.31 & 57 &    & $0.34$ & 0.25 & 53 &     & $0.25$ & 0.16 & 60    \\
\hline
Leo A & 798    & $0.52$ & 0.08 & 58 &    & $0.55$ & 0.08 & 60 &    & $0.41$ & 0.07 & 56 &     & $0.28$ & 0.04 & 62    \\
\hline
GR 8 & 2178    & $0.55$ & 0.32 & 59 &    & $0.58$ & 0.32 & 61 &    & $0.42$ & 0.25 & 56 &     & $0.29$ & 0.16 & 63    \\
\hline
Pegasus dIrr & 920    & $0.55$ & 0.10 & 59 &    & $0.58$ & 0.11 & 61 &    & $0.44$ & 0.09 & 57 &     & $0.29$ & 0.05 & 63    \\
\hline
UGC 9128 & 2291    & $0.59$ & 0.34 & 60 &    & $0.65$ & 0.37 & 63 &    & $0.47$ & 0.29 & 58 &     & $0.32$ & 0.18 & 64    \\
\hline
UGC 4879 & 1361    & $0.60$ & 0.19 & 61 &    & $0.67$ & 0.21 & 63 &    & $0.50$ & 0.17 & 59 &     & $0.32$ & 0.10 & 65    \\
\hline
KK 258 & 2230    & $0.77$ & 0.41 & 66 &    & $0.89$ & 0.45 & 69 &    & $0.68$ & 0.37 & 65 &     & $0.39$ & 0.20 & 68    \\
\hline
UGCA 86 & 2965    & $0.81$ & 0.55 & 67 &    & $0.97$ & 0.63 & 71 &    & $0.74$ & 0.50 & 67 &     & $0.41$ & 0.27 & 69    \\
\hline
\end{tabular}
\end{table}
\clearpage
\begin{table}
\tablewidth{0.97\textwidth}
\centering
\contcaption{Mean number of satellites with $M_* > 10^4$\msun within a $0.56^\circ$ radius field of view around Local Group dwarf field galaxies}
\begin{tabular}{lcccccccccccccccc} 
\hline
\hline
\textbf{Name} &  \boldmath$D_{\sun} \, [\mathrm{kpc}]$ & \multicolumn{3}{c}{\textbf{Moster}} && \multicolumn{3}{c}{\textbf{GK14}} && \multicolumn{3}{c}{\textbf{GK16}} && \multicolumn{3}{c}{\textbf{Brook}} \\
 \cline{3-5}\cline{7-9}\cline{11-13} \cline{15-17} \\ 
  &  & $\bar{N}_{\rm{lum}}$ &  $\bar{N}_{\rm{fov}}$ & $R_{\rm{vir}} \, \rm{[kpc]}$  &  & $\bar{N}_{\rm{lum}}$ & $\bar{N}_{\rm{fov}}$ & $R_{\rm{vir}} \, \rm{[kpc]}$ & & $\bar{N}_{\rm{lum}}$  & $\bar{N}_{\rm{fov}}$ & $R_{\rm{vir}} \, \rm{[kpc]}$ &  & $\bar{N}_{\rm{lum}}$ & $\bar{N}_{\rm{fov}}$ & $R_{\rm{vir}} \, \rm{[kpc]}$   \\
\hline
DDO 99 & 2594    & $0.82$ & 0.49 & 67 &    & $0.98$ & 0.56 & 71 &    & $0.75$ & 0.45 & 67 &     & $0.40$ & 0.23 & 69    \\
\hline
UKS 2323-326 & 2208    & $0.84$ & 0.43 & 67 &    & $0.99$ & 0.48 & 72 &    & $0.77$ & 0.39 & 68 &     & $0.41$ & 0.20 & 70    \\
\hline
UGC 8508 & 2582    & $0.90$ & 0.53 & 69 &    & $1.06$ & 0.58 & 73 &    & $0.82$ & 0.48 & 69 &     & $0.44$ & 0.25 & 71    \\
\hline
KKs 3 & 2120    & $0.97$ & 0.45 & 70 &    & $1.22$ & 0.52 & 76 &    & $0.92$ & 0.42 & 72 &     & $0.46$ & 0.21 & 72    \\
\hline
NGC 4163 & 2858    & $1.17$ & 0.69 & 75 &    & $1.56$ & 0.84 & 82 &    & $1.22$ & 0.70 & 78 &     & $0.56$ & 0.33 & 76    \\
\hline
WLM & 933    & $1.26$ & 0.16 & 77 &    & $1.69$ & 0.18 & 85 &    & $1.32$ & 0.16 & 80 &     & $0.58$ & 0.07 & 77    \\
\hline
Sextans A & 1432    & $1.28$ & 0.32 & 77 &    & $1.70$ & 0.37 & 85 &    & $1.33$ & 0.31 & 81 &     & $0.60$ & 0.15 & 78    \\
\hline
DDO 125 & 2582    & $1.31$ & 0.68 & 78 &    & $1.73$ & 0.81 & 86 &    & $1.41$ & 0.70 & 82 &     & $0.61$ & 0.31 & 78    \\
\hline
DDO 190 & 2793    & $1.37$ & 0.76 & 79 &    & $1.87$ & 0.94 & 87 &    & $1.47$ & 0.77 & 83 &     & $0.63$ & 0.35 & 79    \\
\hline
Sextans B & 1426    & $1.40$ & 0.33 & 79 &    & $1.86$ & 0.38 & 87 &    & $1.48$ & 0.33 & 83 &     & $0.64$ & 0.15 & 79    \\
\hline
IC 3104 & 2270    & $1.49$ & 0.64 & 81 &    & $2.03$ & 0.77 & 90 &    & $1.63$ & 0.65 & 86 &     & $0.67$ & 0.29 & 81    \\
\hline
NGC 3109 & 1300    & $1.65$ & 0.31 & 84 &    & $2.29$ & 0.37 & 93 &    & $1.85$ & 0.32 & 89 &     & $0.75$ & 0.14 & 83    \\
\hline
NGC 6822 & 459    & $1.85$ & 0.05 & 87 &    & $2.61$ & 0.07 & 98 &    & $2.09$ & 0.05 & 94 &     & $0.82$ & 0.02 & 86    \\
\hline
IC 1613 & 755    & $1.84$ & 0.14 & 87 &    & $2.60$ & 0.16 & 98 &    & $2.16$ & 0.14 & 94 &     & $0.81$ & 0.06 & 86    \\
\hline
IC 4662 & 2443    & $2.46$ & 0.95 & 95 &    & $3.61$ & 1.18 & 109 &    & $2.96$ & 1.02 & 105 &     & $1.10$ & 0.43 & 94    \\
\hline
IC 5152 & 1950    & $2.81$ & 0.75 & 100 &    & $4.28$ & 0.94 & 115 &    & $3.74$ & 0.84 & 113 &     & $1.29$ & 0.35 & 99    \\
\hline
\hline
\end{tabular}
\tablecomments{$\bar{N}_{\rm{fov}}$ indicates the mean number of luminous satellites with $M_* > 10^4$\msun within a field of view of radius $0.56^\circ$ (corresponding to a footprint of equal area as the Solo Dwarfs Project) centered on target Local Group dwarf galaxies, as predicted with various AM models. Galaxies with a larger heliocentric distance and smaller AM model inferred virial radius, $R_{\rm{vir}}$, will have a larger fraction of their volume surveyed in the field of view. The total mean number of satellites within each galaxy's virial volume, $\bar{N}_{\rm{lum}}$, is listed for comparison.}
\end{table}